\documentclass[prl,a4,twocolumn, notitlepage,superscriptaddress, nofootinbib, showpacs,tightenlines]{revtex4-1}

\usepackage{array}
\usepackage{color}
\usepackage{multirow}
\usepackage{amsmath} 
\usepackage{amsthm} 
\usepackage{amssymb}	

\usepackage{graphicx}
\usepackage{dcolumn}

\usepackage{kbordermatrix} 
\usepackage{appendix}

\newcommand{\bla}{\color{black}}

\newcommand{\defcol}{\bla}
\newcommand{\ampcol}{\bla}
\newcommand{\concol}{\bla}
\newcommand{\errcol}{\bla}

\newcommand{\Htogcol}{\bla}

\newcommand{\Utot}{U\bla}

\newcommand{\Uc}{\concol U_c\bla}
\newcommand{\UcD}{\concol U_c^\dagger\bla}
\newcommand{\Uerr}{\errcol \tilde{U}\bla}

\newcommand{\Hc}{\concol {H}_c\bla}
\newcommand{\Herr}{\errcol {H}_0\bla}
\newcommand{\Htog}{\Htogcol\tilde{H}_0\bla}

\newcommand{\Hd}{\defcol H_0^{(z)}\bla}
\newcommand{\Ha}{\ampcol H_0^{(\Omega)}\bla}

\newcommand{\Ba}{\ampcol\beta_{_{\Omega}}\bla}
\newcommand{\Bd}{\defcol\beta_z\bla}
\newcommand{\Sa}{\ampcol S_{_{\Omega}}\bla}
\newcommand{\Sd}{\defcol S_z\bla}
\newcommand{\Fa}{\ampcol F_{_{\Omega}}\bla}
\newcommand{\Fd}{\defcol F_z\bla}

\newcommand{\PAL}{\text{PAL}}

\newcommand{\Walsh}{w}

\newcommand{\ampH}{\tilde{X}}
\newcommand{\ampP}{X}
\newcommand{\WAMGRabiPlus}{X_+}
\newcommand{\WAMGRabiMin}{X_-}
\newcommand{\MinDim}{\mathcal{M}}

\newcommand{\EV}{\boldsymbol{a}}

\newcommand{\sigvec}{\boldsymbol{\sigma}}

\newcommand{\sigphiL}{\hat{\sigma}_{\phi_l}}

\newcommand{\Id}{\mathbf{I}}

\newcommand{\rvec}{\vec{\boldsymbol{r}}}

\newcommand{\ket}[1]{\left| #1 \right>} 

\newcommand{\CtrlSpace}{\mathfrak{C}}

\newcommand{\CPSMatElem}{\boldsymbol{\Gamma}}

\newcommand{\phiSK}{\phi_{_\text{SK1}}}

\newcommand{\WAMampP}{X}
\newcommand{\WAMampH}{\tilde{\WAMampP}}
\newcommand{\avec}{\boldsymbol{a}}

\renewcommand{\kbldelim}{[} 
\renewcommand{\kbrdelim}{]}

\begin{document}
\title{Experimental noise filtering by quantum control}%
\author{A. Soare}
\email{\emph{These two authors contributed equally to this work.}}
\author{H. Ball}
\email{\emph{These two authors contributed equally to this work.}}
\affiliation{ARC Centre for Engineered Quantum Systems, School of Physics, The
University of Sydney, NSW 2006 Australia\\National Measurement Institute, West Lindfield, NSW 2070 Australia}

\author{D. Hayes}
\altaffiliation{\emph{Present address:} Lockheed Martin Corporation}
\author{J. Sastrawan}
\author{M. C. Jarratt}
\affiliation{ARC Centre for Engineered Quantum Systems, School of Physics, The
University of Sydney, NSW 2006 Australia\\National Measurement Institute, West Lindfield, NSW 2070 Australia}
\author{J. J. McLoughlin}
\affiliation{ARC Centre for Engineered Quantum Systems, School of Physics, The
University of Sydney, NSW 2006 Australia\\National Measurement Institute, West Lindfield, NSW 2070 Australia}
\author{X. Zhen}
\affiliation{Tsinghua University, Beijing, People's Republic of China}
\author{T. J. Green}
\author{M. J. Biercuk}
\email{michael.biercuk@sydney.edu.au}
\affiliation{ARC Centre for Engineered Quantum Systems, School of Physics, The
University of Sydney, NSW 2006 Australia\\National Measurement Institute, West Lindfield, NSW 2070 Australia}

\date{\today}%


 
\begin{abstract}
Extrinsic interference is routinely faced in systems engineering, and a common solution is to rely on a broad class of \emph{filtering} techniques in order to afford stability to intrinsically unstable systems or isolate particular signals from a noisy background.  For instance, electronic systems are frequently designed to incorporate electrical filters composed of, \emph{e.g.} RLC components~\cite{Pozar}, in order to suppress the effects of out-of-band fluctuations that interfere with desired performance.  Experimentalists leading the development of a new generation of quantum enabled technologies similarly encounter time-varying noise in realistic laboratory settings.  They face substantial challenges in either suppressing such noise for high-fidelity quantum operations~\cite{Schmidt2005, Foletti_NP2009, Britton2012, JessenPRL2013, Islam03052013} or controllably exploiting it in quantum-enhanced sensing~\cite{Lukin2013, Hollenberg_NP_2011, Lukin_Nature2013, RBMagnetometry, Cooper14} or system identification tasks~\cite{Bylander2011}, due to a lack of efficient, validated approaches to understanding quantum dynamics in the presence of time-varying noise.  In this work we use the theory of quantum control engineering~\cite{Tarn2003, James2007} and experiments with trapped $^{171}$Yb$^{+}$ ions to construct novel \emph{noise spectral filters} with user-defined properties and compatible with incorporation into arbitrary control operations.    Our results provide the first experimental validation of generalized filter-transfer functions for arbitrary quantum control operations~\cite{GreenPRL2012, GreenNJP2013}, and demonstrate their utility for developing novel robust control and sensing protocols.  We provide a detailed framework for filter synthesis appropriate for arbitrary single-qubit state transformations and experimentally validate the performance of the resulting noise filters.  These experiments provide a significant advance in the theory of quantum control and unlock new capabilities for the emerging field of quantum systems engineering.
\end{abstract}

\maketitle

\begin{figure*}[tp]
\includegraphics[width=14cm]{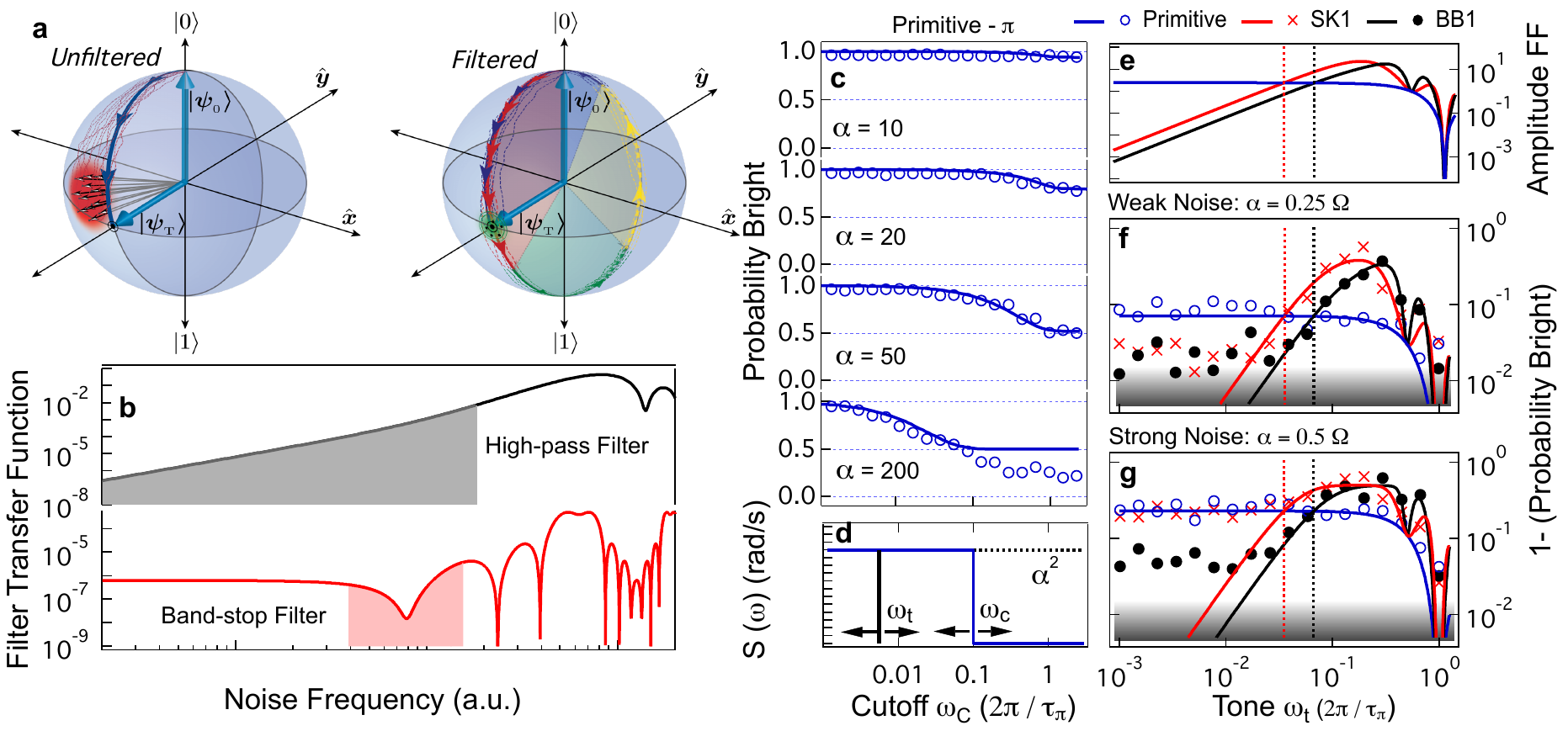}
\caption{\label{Fig:FFvalidation} Noise filters and experimental validation of the predictive power of the filter transfer function.  a) Time-varying noise during an operation (a rotation on the Bloch sphere, here $\theta=\pi/2$, $\phi=0$ ) produces a broad range of outcomes (red uncertainty cone, left) and may yield an offset of the average final state from the target state, measured as operational infidelity.  Filtered trajectory, depicted as a user-defined modulation pattern on the control (colored segments), changes the measured fidelity by reducing the uncertainty due to noise in a specified band.  b) Schematic representation of noise filters of interest - shaded areas represent filter stop-bands - crafted by control modulation as indicated above.  c) Measurements of operational fidelity with engineered dephasing noise for primitive $\pi$ rotation, $\ket{0}\to\ket{1}$, as a function of dimensionless noise cutoff frequency overlaid with FF-based calculations of $\mathcal{F}_{\chi}(\tau)$.   Decay to value 0.5 corresponds to full decoherence.  For $\omega_{c}/2\pi=1$ the highest frequency contribution to $S_{z}(\omega)$ undergoes a complete cycle of oscillation over $\tau_{\pi}$, indicating that the noise is time-dependent on the scale of a single experiment even for $\omega_{c}/2\pi\ll1$. Each data point is the result of averaging over 50 different noise realizations.   (d) Schematic representation of the quasi-white noise power spectrum employed in (c) and single-tone power spectrum employed in (f-g).  Noise strength parameterized by $\alpha$.  e) Calculated $F_{\Omega}(\omega)$, for primitive and compensating $\pi$ pulses (see~\cite{MerrillArXv2012}).  Vertical lines indicate frequencies where $F_{\Omega}(\omega)$ for SK1 (red) and BB1 (black) cross primitive (blue), indicating an expected inversion of performance. f)-g) Swept-tone multiplicative amplitude noise measurements , $S_{\Omega}(\omega)\propto\delta(\omega_{t}-\omega)$, for various $\pi$ rotations (averaged over 20 noise realizations).  Vertical axis is a proxy for measured operational \emph{infidelity}.  Solid lines indicate $1- \mathcal{F}_{\chi}(\tau)$, revealing good agreement in the weak noise limit (f) across three decades of frequency, down to measurement fidelity limit, $\sim98.5\%$, indicated by grey shading.  Measured gate-error crossover points correspond well with crossovers in the FFs for these gates (vertical dashed lines).  Detailed performance differences between protocols in the low-error limit can be revealed through randomized benchmarking, as performed later (Fig.~\ref{Fig:F2}).   g) Strong-error limit, first-order approximations are violated and contributions from higher-order Magnus terms contribute to the measured error in the low-frequency limit, yielding (expected) differences between SK1 and BB1 not captured by the FF. }
\end{figure*}

The presence of time-dependent noise in either the environment or control of a single qubit randomizes the intended evolution trajectory of a state $\ket{\psi_{0}}\to\ket{\psi_{\text{T}}}$ (Fig.~\ref{Fig:FFvalidation}a), ultimately reducing the fidelity of the operation.  This is captured as $\mathcal{F}_{\chi}(\tau) = \frac{1}{2}(1 + e^{-\chi(\tau)})$, where $\chi(\tau)=\frac{1}{\pi}\sum_i\int_{0}^{\infty}d\omega S_{i}(\omega)F_{i}(\omega)$, and $\tau$ is the total duration of the operation.  In this expression for fidelity, the integral considers contributions from independent noise processes through their frequency-domain power spectra $S_{i}(\omega)$, $i\in\{z,\Omega\}$, capturing dephasing along $\hat{z}$ and amplitude noise co-rotating with a resonant drive field (see \emph{Supplementary Material}).
 
The quantities $F_{i}(\omega)$ describe the effective frequency response of any applied control operation (including free evolution),  and are referred to as filter transfer functions, ``FF''s for the control, providing an efficient manner to understand the dynamical response of a controlled quantum system in a time-varying environment~\cite{Note1, KurizkiPRL2001, KurizkiPRL2004,UhrigPRL2007, CywinskiPRB2008, BiercukJPB2011}.  Despite their appeal -- FFs may be characterized using a standard engineering approach considering frequency passbands, stopbands, and filter order -- to date, the only demonstrations of FFs in quantum control have been undertaken in the simple case of application of the \emph{identity} operator (dynamical decoupling)~\cite{BiercukNature2009, Suter_PRL2011, Bylander2011}.  In this case the analytics are dramatically simplified due to assumptions of unbounded ``bang-bang'' control.  

More generally, calculating $F_{i}(\omega)$ and hence characterizing the spectral response of an \emph{arbitrary} bounded-strength control operation is challenging in cases where the noise and control Hamiltonians do not commute (e.g. a driven $\sigma_{x}$ rotation in the presence of $\sigma_{z}$ dephasing).  With recent theoretical developments FFs may now be calculated analytically for arbitrary control~\cite{GreenPRL2012, GreenNJP2013}; it is this more general case where the impact of noise filtering and the filter transfer functions may have the most significant impact on the quantum engineering community, and where experimental tests are most germane.

To see this we may consider the various tasks that might be of interest to an experimentalist engaged in quantum engineering and the role of noise spectral filtering in these applications.   Noise filtering itself is achieved through construction of a control protocol (Fig.~\ref{Fig:FFvalidation}a) which reduces the \emph{controllability} of the quantum system by the noisy environment \emph{over a defined frequency band} by suitably modifying $F_{i}(\omega)$.  In quantum information an experimentalist may aim to suppress broadband low-frequency noise in order to maximize the fidelity of a bounded-strength quantum logic operation (Fig.~\ref{Fig:FFvalidation}b, upper trace).   Alternatively, in quantum enabled sensing or system identification he or she may perform narrowband spectral characterization of a given operation (Fig.~\ref{Fig:FFvalidation}b, lower trace), where the measured infidelity under filter application represents the signal of interest~\cite{Bylander2011, Cooper14}.  The FFs are simple analytic objects which enable the extraction of exactly this information calculated for arbitrary control and arbitrary universal noise.

In our experimental system, based on the 12.6 GHz qubit transition in $^{171}$Yb$^{+}$ (see \emph{Supplementary Material}), we are able to perform quantitative tests of operational fidelity revealing the spectral characteristics of arbitrary control operations; these may then be compared against calculations of $\mathcal{F}_{\chi}(\tau)$ as a fundamental test of FF validity.  A key tool in our studies is bath engineering~\cite{SoareBath}, in which we add noise with user-defined spectral characteristics to the control system, producing well controlled unitary dephasing or depolarization.

In our first experiment we measure the probability that a $\pi_{x}$-pulse drives qubit population from the dark state to the bright state, $\ket{0}\to\ket{1}$, while varying the high-frequency cutoff, $\omega_{c}$, of an engineered non-Markovian dephasing bath (Fig.~\ref{Fig:FFvalidation}c).  As the high-frequency cutoff of the noise is increased and fluctuations fast compared to the control ($\tau_{\pi}$) are added to the noise power spectrum, $S_{z}(\omega)$, errors accumulate reducing the measured fidelity.  We calculate $\mathcal{F}_{\chi}(\tau)$ using the form of the noise and the analytic FF for a driven primitive gate under dephasing~\cite{GreenNJP2013}, finding good agreement with experimental measurements using \emph{no free parameters}, thus verifying the predictive power of the FF.   

The FFs for much more complex control such as compensating composite pulses~\cite{Vandersypen2004,MerrillArXv2012} can be calculated as well~\cite{Kabytayev2014}, revealing their sensitivity to time-dependent noise - an important characteristic for deployment in quantum information settings.   We experimentally demonstrate a form of quantum system identification, reconstructing the amplitude-noise filter functions, $F_{\Omega}(\omega)$, for SK1 and BB1 $\pi_{x}$-pulse sequences by measuring operational fidelity in the presence of a swept narrowband amplitude noise signal~\cite{SoareBath}.  The resulting measured (in)fidelity effectively traces out the FF; key features such as performance-crossover frequencies between primitive and compensating gates and deep notches in the filter at high frequency are quantitatively reproduced in experimental measurements.  Again, first-order fidelity calculations, $\mathcal{F}_{\chi}(\tau)$, match data well in the weak noise limit (Fig.~\ref{Fig:FFvalidation}f) with \emph{no free parameters}.  

 Ultimately, the underlying physical principles giving rise to the analytic form of $F_{i}(\omega)$ are based on the well tested average Hamiltonian theory~\cite{Magnus1954, Waugh1968} common to NMR and quantum information.  Despite this shared theoretical foundation, the calculation of spectral filtering properties is quite distinct from finding compensating-pulse protocols giving error-compensation in a Magnus expansion.  This distinction is important as time-varying colored classical noise is commonly encountered in laboratory settings, but high-Magnus-order composite pulses need not be efficient noise spectral filters (see \emph{Supplementary Material} and ~\cite{Kabytayev2014}).  
  
Measurements of SK1 and BB1 fidelity highlight this difference.  Both gates provide similar filtering of time-dependent noise, given by the \emph{filter order},~Fig.~\ref{Fig:FFvalidation}e, despite the significant differences in their construction; BB1 is designed to suppress higher-order Magnus terms than SK1 (see \emph{Supplementary Material}).  In the strong-error limit (Fig.~\ref{Fig:FFvalidation}g), however, significant performance deviations arise at low frequencies; the increased Magnus-order cancellation of BB1 improves quasi-static error suppression relative to SK1 (the regime for which both sequences were originally crafted).  Overall, frequency-domain characteristics are captured accurately through the FF in the weak noise limit (Fig.~\ref{Fig:FFvalidation}f), but would not be directly evident through a Magnus expansion.  


\begin{figure}[bp]
\includegraphics[width=7.5cm]{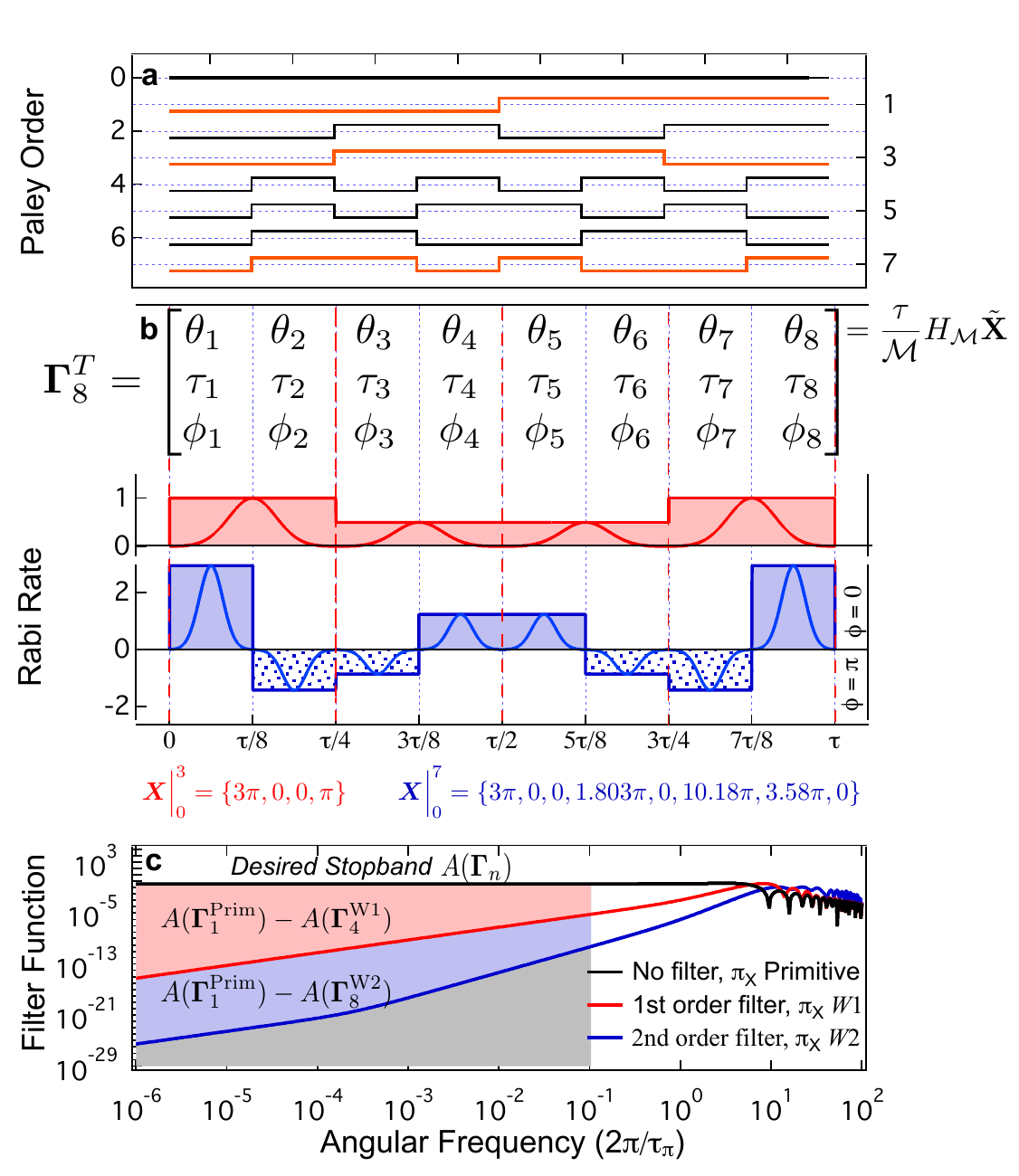}
\caption{\label{Fig:FilterConstruction} Synthesis of high-pass amplitude-modulated filters from the Walsh functions.   a) The first eight Walsh functions used in filter synthesis, $\{\PAL_{0}, \PAL_{7}\}$, with maximum-Hamming-weight-indexed functions highlighted.  b) Representative amplitude profiles for filter constructions found via numerical search over the Walsh basis with four (red, denoted $W1$) and eight (blue, denoted $W2$) time steps.  Vertical axis represents $\Omega$, the Rabi rate per time step; negative values indicate $\pi$-phase shifts.  Synthesis may be performed over square (flat-top) pulse-segments or Gaussian-shaped pulse segments with results differing only in the resulting Walsh coefficients.  The matrix representing filter characteristics over eight segments is superimposed on the amplitude profiles (for $n=4$, neighboring segments between red dashed lines are combined).  The first row (the angles of rotation in each segment of the filter) is determined via Walsh synthesis, indicated by the vectors $\boldsymbol{X}^T\Big|_{0}^{n}$, containing the spectral weights over $\PAL_0 \to \PAL_n$.  In the case of Gaussian pulse envelopes Walsh synthesis sets the first line, $\theta_{l}$.  The symbol $\boldsymbol{\tilde{X}}$ indicates reordering for Hadamard synthesis, with listed coefficients appropriate for square pulse envelopes.  c) The filter transfer function for a primitive $\pi_{x}$ rotation and for synthesized noise filters.  Performance improvement over the desired stopband of the filter captured in cost function $A(\CPSMatElem_{4(8)}^{W1(W2)})$ and its difference relative to that for the primitive operation, $A(\CPSMatElem_{1}^{\text{Prim}})$.  Filter $W1$ gives improvement indicated by the red shading, with additional improvement in the cost function given by $W2$ indicated by blue shading.  
}
\end{figure}

These simple but powerful validations of the generalized filter transfer function's predictive power now open the possibility of demonstrating the \emph{construction} of noise filters with a specified spectral response, employing the filter transfer functions as key analytic tools. Filters may take a wide variety of forms - including high-pass filters for broadband noise suppression and band-stop filters useful for narrowband noise characterization (Fig.~\ref{Fig:FFvalidation}b).

 In the example that follows, we focus on a common setting in which we aim to improve operational fidelity by reducing the influence of broadband non-Markovian noise on a target state transformation.  Filters are realized as sequences of time-domain control operations with tunable pulse amplitude and phase, similar in spirit to compensating composite pulses in NMR~\cite{Vandersypen2004,MerrillArXv2012,Kabytayev2014}, dynamically corrected gates (DCGs) in quantum information~\cite{khodjasteh2009dcg, DasSarmaGate}, and open-loop modulated pulses in quantum control~\cite{UhrigPRA2012, Altafini2013}.  In this setting we wish to synthesize a filter with arbitrary, user-defined spectral characteristics captured by a cost-function, $A(\CPSMatElem_{n})$, to be minimized over $n$ pulse segments in a filter construction. 
  
With the unique task of creating filters tailored to a given noise spectrum in hand, we introduce a basic framework for filter construction leveraging the filter-transfer function.  An arbitrary $n$-segment filter is represented over successive timesteps through the matrix quantity $\CPSMatElem_n (\theta_{l},\tau_{l}, \phi_{l})$ (Fig.~\ref{Fig:FilterConstruction}b, \emph{Supplementary Material}) describing the properties of a near-resonant carrier frequency enacting driven operations. In each segment of duration $\tau_{l}$ we perform a driven operation generating a rotation through an angle $\theta_l = \int_{t_{l-1}}^{t_{l}}\Omega_l(t)dt$ about the axis $\rvec_l = (\cos(\phi_l),\sin(\phi_l),0)$, with $\Omega_l(t)$ the Rabi rate over the $l$th pulse segment.  


\begin{figure*}[t]
\includegraphics[width=14cm]{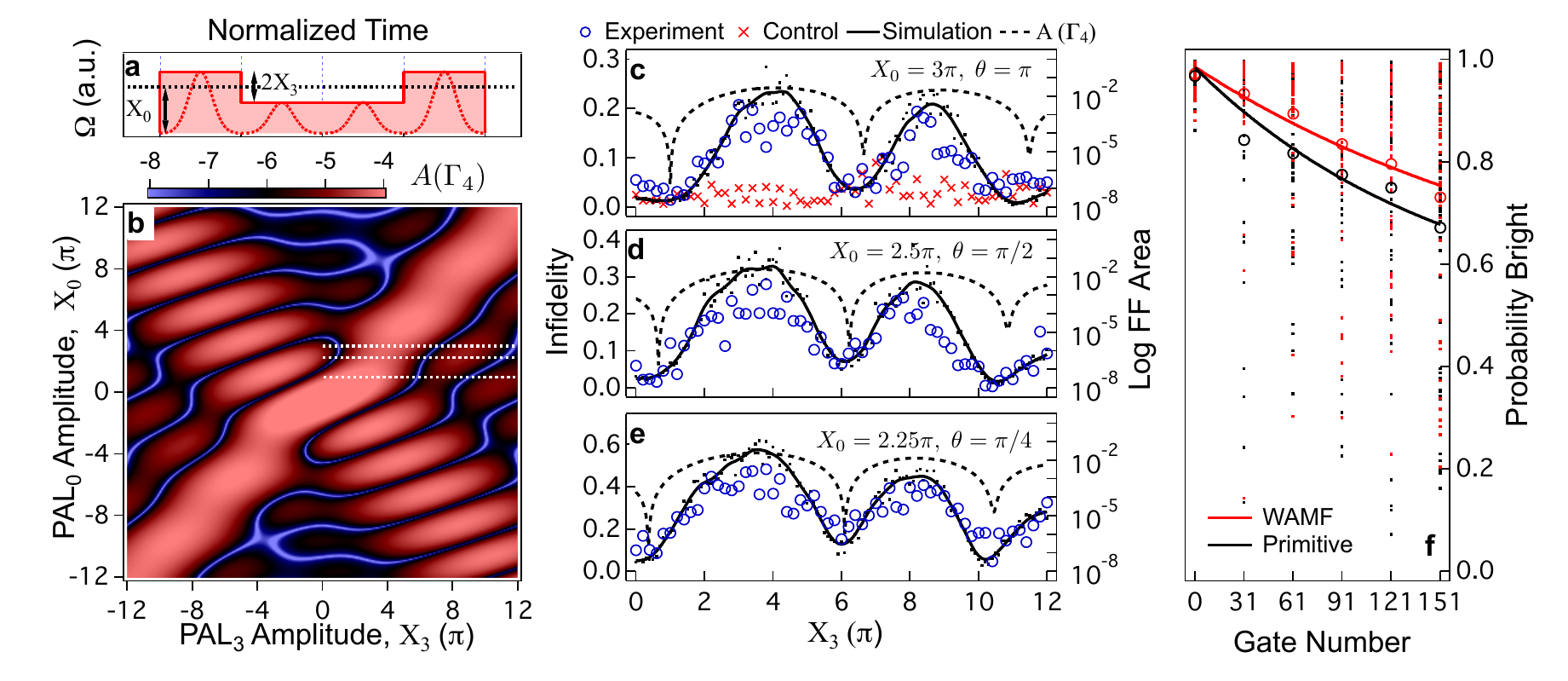}
\caption{\label{Fig:F2}Construction of the first-order Walsh amplitude modulated dephasing-suppressing filter.   a) Schematic representation of Walsh synthesis for a four-segment amplitude-modulated filter (WAMF).  Walsh synthesis can be used to determine either the modulating envelope of square pulse segments, or the net area of discrete Gaussian pulses with differing amplitudes.  b) Two-dimensional representation of the integral metric defining our target cost function, $A(\CPSMatElem_{4})$ integrated over the stopband $\omega\in [10^{-9}, 10^{-1}]\tau^{-1}$. Areas in blue minimize $A(\CPSMatElem_{4})$, representing effective filter constructions. The $X_{0}$ determines the net rotation enacted in a gate while $X_{3}$ determines the modulation depth, as represented in a).   White lines indicate possible constructions for filters implementing rotations of $\theta=\pi$, $\theta=\pi/2$, and $\theta=\pi/4$ from top to bottom. c)-e) Experimental measurement of gate infidelity (left axis) for rotations constructed from various Walsh coefficients in the presence of engineered noise ($\omega_{c}/2\pi=20$ Hz).  Black dots and line represented calculated fidelity by Schroedinger equation integration (raw and smoothed respectively).  All values of $\ampP_3$ for a given $\ampP_{0}$ implement the same net rotation, indicated by control experiment with no noise. Total rotation time is scaled with $\ampP_{3}$ to preserve a maximum \emph{Rabi rate}.  Black dashed line (right axis) corresponds to $A(\CPSMatElem_{4})$ from panel (a).   In experiments we always perform a net $\pi$ rotation $\ket{0}\to\ket{1}$ by sequentially performing identical copies of rotations for $\theta<\pi$.  f) Randomized benchmarking results (50 randomizations) demonstrating superior performance of modulated gate in the small-error limit (infidelity $<0.5\%$ per gate), see \emph{Supplementary Material}.}  
\end{figure*}

To provide \emph{efficient} solutions to filter design we restrict our control space and focus on constructions synthesized using concepts from functional analysis in the basis set of \emph{Walsh functions} - square-wave analogues of the sines and cosines~\cite{Walsh_Beauchamp, HayesPRA2011, Cooper14} (Fig.~\ref{Fig:FilterConstruction}a).  This approach is by no means the only basis set for composite gate construction~\cite{OwrutskyArXv2012,JonesNJP2012}, but provides significant benefits for our problem~\cite{HayesPRA2011}.  For instance, their piecewise-constant construction builds intrinsic compatibility with discrete clocking~\cite{Hodgson2010} and classical digital logic, while the well characterized mathematical properties of the Walsh functions provide a basis for establishing simple \emph{analytic} filter-design rules, and flexibility in realizing a wide variety of filter forms (see \emph{Supplementary Material}).   

As an example we synthesize filters via weighted linear combination of Paley-ordered Walsh functions, $\PAL_k(x)$, designed to suppress time-varying dephasing noise over a low-frequency stopband while implementing a bounded-strength driven rotation about the $x$-axis on the Bloch sphere.  In this case the Walsh-synthesized waveform dictates an amplitude modulation pattern for the control field over discrete time-segments.  Importantly, Walsh filter synthesis is compatible with pulse segments possessing \emph{arbitrary pulse envelope}, including sequences of \emph{e.g.} square (used here) or Gaussian pulse segments (Fig.~\ref{Fig:FilterConstruction}b).

Walsh-synthesis design rules dictate that we implement our filtered rotation by $\theta_{x}$ over a minimum of four discrete steps, permitting synthesis over $\PAL_0$ to $\PAL_3$.  Within this small set, the coefficient of $\PAL_0$, denoted $\ampP_{0}$, sets the total rotation angle $\theta\mod{2\pi}$ for the modulated driven evolution, and only nonzero $\ampP_{3}$ preserves symmetry.  Relative to an unfiltered primitive gate, $F_{z}(\omega)$ for a construction giving effective filtered evolution, $W1$, shows increased steepness in the stopband (Fig.~\ref{Fig:FilterConstruction}c, red), reducing $A(\CPSMatElem_{4})$ (here, $\theta=\pi)$.  Doubling the number of pulse-segments, and allowing Walsh synthesis over coefficients $\{\ampP_{0},...,\ampP_{7}\}$ yields filter $W2$, which increases the order of time-domain noise filtering (steeper rolloff), further reducing the cost function for optimization (blue shaded area in Fig.~\ref{Fig:FilterConstruction}c).   Interestingly, $W1$ is a special case of an analytically constructed dynamically corrected NOT gate (a $\pi$-rotation)~\cite{khodjasteh2009dcg}.  

Filters $W1$ and $W2$ were found using a Nelder-Mead simplex optimization over Walsh coefficients and are representative, rather than unique solutions.  In Fig.~\ref{Fig:F2}b we show the calculated cost function, $A(\CPSMatElem_{4})$, as a function of $\ampP_{0}$ and $\ampP_{3}$.  Blue areas meet our minimized target indicating useful filters, revealing a wide variety of possible constructions with favorable filtering characteristics.  


We experimentally test the performance of four-segment amplitude-modulated filters by scanning over $\ampP_{3}$ for fixed $\ampP_{0}$ (denoted by white dotted lines in Fig.~\ref{Fig:F2}b).  Values of $\ampP_{3}$ minimizing $A(\CPSMatElem_{4})$ (dips in the dashed trace, right axis) also minimize the experimentally measured infidelity in the presence of engineered low-frequency noise (open circles, left axis).  This behavior is observed for various target rotation angles of interest (Fig.~\ref{Fig:F2}c-e), with predicted shifts in the optimal values of $\ampP_{3}$ with changes in $\ampP_{0}$ borne out through experiment.  In the small error limit we also show that a filtered $\pi$ pulse ($W1$) outperforms the best primitive operation via randomized benchmarking (Fig.~\ref{Fig:F2}f), despite being four times longer.   These measurements thus validate the concept of spectral filter synthesis based on analytic FFs.


In summary, our experiments have shown that filter transfer functions are effective tools for the characterization of complex time-domain control operations in the presence of universal noise, revealing fine details about their spectral features.  We have provided the first experimental validation of their predictive power for arbitrary quantum logic, observing good agreement between measurements on trapped ions and FF-calculated fidelity using \emph{no free parameters}.   In addition, we have shown how one may leverage these analytic constructs for the synthesis of a wide range of time-domain noise filters with user-specified characteristics from the Walsh basis functions. 


Our focus has been on providing a validated framework for the vital task of filter construction to complement existing techniques, rather than attempting to find maximum-performance error-robust gates. Our results on high-pass noise filters, for instance, add to existing compensating pulse sequences designed for quasi-static noise, as well as gate constructions with interleaved dynamical decoupling that seek to periodically ``refocus'' a quantum trajectory~\cite{Barthel2010, YacobyInterleaved, RBInterleaved, Suter_Interleaved, HansonInterleaved}.  Other applications of this approach include crafting notch filters for noise characterization tasks and extension to the filtering of noise in multiqubit logical operations~\cite{HayesPRL}.  We believe that this simple framework will provide a straightforward path for experimentalists to characterize and suppress the effects of noise in generic quantum coherent systems, ultimately enabling a new generation of engineered quantum technologies.

\begin{acknowledgments}
We thank P. Fisk, M. Lawn, M. Wouters, and B. Warrington for technical assistance and K. Brown, J.T. Merrill and L. Viola for useful discussions.  This work partially supported by the US Army Research Office under Contract Number  W911NF-11-1-0068, and the Australian Research Council Centre of Excellence for Engineered Quantum Systems CE110001013, the Office of the Director of National Intelligence (ODNI), Intelligence Advanced Research Projects Activity (IARPA), through the Army Research Office, and the Lockheed Martin Corporation. 
\end{acknowledgments}



\bibliography{NoiseEngineeringBib}


\newpage
\appendix
\section{Supplementary Material}

We consider a model quantum system consisting of an ensemble of identically prepared noninteracting qubits immersed in a weakly interacting noise bath and driven by an external control device. Working in the interaction picture with respect to the qubit splitting $\omega_a$ state transformations are represented as unitary rotations of the Bloch vector. The generalized time-dependent Hamiltonian is then written
\begin{equation}\label{Eq:ham0}
H(t)=\Hc(t)+ \Herr(t)
\end{equation}
where $\Hc(t)$ describes perfect control of the qubit state, e.g. via an ideal external driving field, and the noise Hamiltonian $\Herr(t)$ captures undesirable interactions with a (universal) noise bath. 


The specific forms taken by $\Hc(t)$ and $H_0(t)$ in this work are given in the sections below, where we treat both dephasing (detuning) and amplitude-damping (coherent relaxation) noise processes. We will begin with this model to craft time-dependent noise filters, and detail this method in the following sections.


\subsection{Defining the control space}

Representing the qubit state on the Bloch sphere, state manipulation maps to a rotation of the Bloch vector in $\mathbb{R}^3$ and described by the unitary $U(\theta, \hat{\sigma}_{\rvec}): = \exp\big(\frac{-i\sigvec\cdot\rvec\theta}{2}\big)$, reflecting the homeomorphism between $SU(2)$ and $SO(3)$. In effect, the spin operator $\hat{\sigma}_{\rvec} := \rvec\cdot\sigvec$ generates a rotation though an angle $\theta$ about an axis defined by the unit vector $\rvec\in\mathbb{R}^3$. For our purpose control takes the form of a composite pulse sequence consisting of $n$ such unitaries executed over a time period $[0, \tau]$, with the $l$th pulse in the sequence written  
\begin{align}
\label{PrimitivePulseForm}P_l &:= U(\theta_l,\sigphiL) = \exp\Big[-i\frac{\sigphiL}{2}\int_{t_{l-1}}^{t_{l}}\Omega_{l}(t) dt\Big]\\
\label{SpinOperatorDefinition}\sigphiL&:= \cos(\phi_l)\hat{\sigma}_x+\sin(\phi_l)\hat{\sigma}_y.
\end{align}
Here $\Omega_l(t)$ is the Rabi rate with arbitrary amplitude envelope in a single pulse, $\tau_l=t_{l}-t_{l-1}$ is the pulse duration, and the spin operator $\sigphiL$, parametrized by $\phi_l\in[0,2\pi]$, generates a rotation $\theta_l = \int_{t_{l-1}}^{t_{l}} \Omega_{l}(t)dt$ of the Bloch vector about an axis $\rvec_l \equiv (\cos(\phi_l),\sin(\phi_l),0)$ in the $xy$-plane\footnote{For a resonantly driven qubit $\phi_l$ is the phase of the driving field and $\Omega_l$ is linearly proportional to the driving amplitude. }. This sequence of control unitaries implies a natural partition of the total sequence duration  $\tau$ into $n$ subintervals $I_l = [t_{l-1},t_l]$, $l\in\{1,n\}$, such that the $l$th pulse has duration $\tau_l = t_l-t_{l-1}$ with $t_{l-1}$ and $t_l$ the start and end times respectively. Here $t_0 \equiv 0$ and $t_n \equiv \tau$. 
The control Hamiltonian associated with this composite pulse sequence takes the form  
\begin{align}\label{ControlHamiltonian}
H_c(t) = \sum_{l=1}^{n} G^{(l)}(t) 
\frac{\Omega_{l}(t)}{2} 
\sigphiL
\end{align}
where the function $G^{(l)}(t)$ is 1 if $t\in I_l$ and zero otherwise. The sequence of $n$ triples $\{(\theta_l,\tau_l,\phi_l)\}_{l=1}^n$ completely characterizes the net effect of the applied control ($P_l = P_l(\theta_l,\Omega_l(t),\tau_l,\phi_l)$) at the end of successive pulse applications.  We define the $n\times3$ \emph{composite pulse sequence} matrix   
\begin{align}\label{TemplateCPSMatrix}
\CPSMatElem_n\hspace{0.25cm}=\hspace{0.25cm}
\begingroup
  \renewcommand*{\arraystretch}{1.25}%
  \kbordermatrix{
           & \theta_l           & \tau_l           & \phi_l     \cr
    P_1    & \theta_1      & \tau_1        & \phi_1 \cr
    P_2    & \theta_2      & \tau_2        & \phi_2 \cr
    \vdots & \vdots        & \vdots        &  \vdots  \cr
    P_n    & \theta_n     & \tau_n         & \phi_n \cr
  }%
\endgroup
\end{align}

\noindent to compactly describe any arbitrary $n$-pulse control sequence. The entire space of such control forms therefore corresponds to an infinite set of $\CPSMatElem_n$ matrices ranging continously over all possible values taken by the control parameters. We denote this set by $\CtrlSpace_n$ and refer to it as the \emph{$n$-pulse control space}. Written formally 
\begin{align*}
\CtrlSpace_n:= \Big\{\CPSMatElem_n\big|\theta_l,\tau_l>0,\hspace{0.2cm} \phi_l\in[0,2\pi],\hspace{0.2cm}l\in\{1,...,n\},\hspace{0.2cm} \Sigma_l^n\tau_l = \tau\Big\}.
\end{align*}


\subsection{Noise bath model}

We consider semi-classical time-dependent dephasing (detuning) and amplitude damping (relaxation) processes, captured respectively through the appearance of stochastic rotations about $\hat{\sigma}_{z}$ and $\hat{\sigma}_\phi:=\cos(\phi)\hat{\sigma}_x+\sin(\phi)\hat{\sigma}_y$. The universal noise Hamiltonian then takes the form $\Herr(t) = \Hd(t) + \Ha(t)$ where $\Hd(t)$ and $\Ha(t)$ denote noise interactions associated with the dephasing and amplitude noise quadratures respectively. The dephasing noise Hamiltonian is then given by 
\begin{align}\label{DephasingNoiseHamiltonian}
\Hd(t) = \Bd(t)\hat{\sigma}_z
\end{align}
where $\Bd(t)$ is a classical stochastic process.  During each pulse we also make the substitution $\Omega_l(t)\longrightarrow(\Omega_l(t)+\Ba(t)\Omega^{(\text{max})}_{l})$ where $\Ba(t)$ describes captures a (multiplicative) stochastic noise scaled by the maximum Rabi rate in a pulse segment in the amplitude of the driving field. Thus the amplitude noise Hamiltonian takes the form
\begin{align}\label{AmplitudeNoiseHamiltonian}
\Ha &= \Ba(t)\sum_{l=1}^{n} G^{(l)}(t) \frac{\Omega^{(\text{max})}_l}{2} \sigphiL
\end{align}
and generates errors in intended rotation angle coaxial with the target rotation axis $\sigphiL$.


\subsection{Calculating operational fidelity \& the first-order approximation}\label{Sec:ErrorEstimation}

In the absence of noise, state evolution is determined by $i\dot{U}_c(t)=\Hc(t)U_c(t)$ with $U_c(t)$ describing a target operation. Including the effects of noise, however, the actual evolution operator $U(t)$ satisfies $i\dot{U}(t) = (\Hc(t)+\Herr(t))U(t)$. A measure of the average gate fidelity $\mathcal{F}_{av}(\tau) = \frac{1}{4}\langle |\text{Tr}(\UcD(\tau)U(\tau))|^2\rangle$ is then obtained using the Hilber-Schmidt inner product, effectively measuring the overlap between the intended and realized operators. These evolution dynamics however are challenging to compute due to sequential application of noncommuting, time-dependent operations giving rise to both dephasing and depolarization errors. Our approach follows the method developed by Green \emph{et al.} ~\cite{GreenNJP2013}, using average Hamiltonian theory and the generalized filter-transfer function formalism to obtain a first order approximation for gate infidelity.


\subsubsection{Magnus Expansion}\label{Magnus Expansion}

We write the total evolution operator $\Utot(t) = \Uc(t)\Uerr(t)$, where the \emph{error propagator} $\Uerr(t)$ satisfies the Schrodinger equation $i\dot{\Uerr}(t) = \Htog(t)\Uerr(t)$ in a frame co-rotating with the control defined by the \emph{Toggling frame Hamiltonian} $\Htog(t) :=\UcD(t)\Herr(t)\Uc(t)$. Thus, in the event that $\Uerr(\tau) = \Id$, the realized evolution operator $U(\tau)$ approaches the target operation $U_c(\tau)$ and errors do not affect the gate. This can be systematized by writing $\Uerr(\tau) = \exp[-i\Phi(\tau)]$ in terms of a time-independent effective error operator $\Phi(\tau) = \sum_{\mu = 1}^\infty\Phi_\mu(\tau)$ with Magnus expansion terms 
\begin{equation*}\label{MagnusExpansionTerms}
\begin{aligned}
\Phi_1(\tau) &= \int_0^\tau dt \Htog(t)\\
\Phi_2(\tau) &= -\frac{i}{2}\int_0^\tau dt_1\int_0^{t_1} dt_2 \big[\Htog(t_1),\Htog(t_2)\big]\\
\Phi_3(\tau) &= \frac{1}{6}\int_0^\tau dt_1\int_0^{t_1} dt_2 \int_0^{t_2} dt_3
\Big\{\Big[\Htog(t_1),\big[\Htog(t_2),\Htog(t_3)\big] \Big]\\&+\Big[\Htog(t_3),\big[\Htog(t_2),\Htog(t_1)\big] \Big]\Big\}\\
&...
\end{aligned}
\end{equation*}
generally taking the form of time-ordered integrals over nested commutators in $\Htog(t)$. 

One may then use vector identities (given Unitary processes) to re-express the so-called \emph{error vector} in the toggling frame, again in an infinite power-series expansion
\begin{eqnarray}\label{Eq:errser}
\boldsymbol{a}(\tau)=\sum_{\mu}^{\infty}\boldsymbol{a}_{\mu}(\tau).
\end{eqnarray}
\noindent This is a reexpression of the Magnus expansion error terms in the language of our control Hamiltonian.  For details of this derivation and the definition of all terms see~\cite{GreenNJP2013}.



Our theoretical predictions based on the filter-transfer function formalism involve an approximation to the \emph{trace} or \emph{gate fidelity} defined by 
\begin{align}\label{Eq:TraceFidelityDefinition}
\mathcal{F}_{av}(\tau) 
=  \frac{1}{4}\langle |\text{Tr}(\Uerr(\tau))|^2\rangle= \frac{1}{4}\langle |\text{Tr}(e^{-i\avec(\tau)\sigvec}|^2\rangle 
\end{align}
where $\avec(\tau)=\sum_{\mu}^{\infty}\avec_{\mu}(\tau)$ is the error vector given in terms of the Magnus expansion. Following the method developed in Ref.  ~\cite{GreenNJP2013} and expanding the exponential in Eq. \ref{Eq:TraceFidelityDefinition} we obtain
\begin{align}\label{TraceFidelityCosExpansion}
\mathcal{F}_{av}(\tau) &= \frac{1}{2}[\langle \cos(2a)\rangle+1]\\
& =  \frac{1}{2}
\Big[
1+\sum_{m=0}^\infty(-1)^m\frac{2^{2m}}{(2m)!}\langle a^{2}\rangle^m
\Big]\\
\langle a^2\rangle &=\sum_{\mu\nu}[
\langle a_1^2\rangle+
\langle a_2^2\rangle+\text{É}\\
&+
2(\langle\avec_1\avec_2^T\rangle+
\langle\avec_1\avec_3^T\rangle+
\langle\avec_2\avec_3^T\rangle+\text{...})
] 
\end{align}
with $a^2 \equiv \avec(\tau)\avec(\tau)^T$ the norm square of the error vector. The full expansion rapidly becomes too complex to write explicitely, however it is convenient to write $\mathcal{F}_{av}=\sum_{k=0}^{\infty}\mathcal{O}(\xi^{2k})$ where $\xi$ is the \emph{smallness parameter} quantifying the RMS deviation of the noise integrated over the sequence duration, $\xi\equiv\Delta\beta\tau/2$~\cite{GreenPRL2012}.  For this series to formally converge we require $\xi^{2}<1$ (see main text and Fig.~\ref{Fig:Pi_FFvalidation}).

Here odd powers of $\xi$ are omitted since these involve ensemble averages over odd powers of the noise strength and vanish under our assumption of zero-mean, Guassian-distributed random variables. Writing the $\mathcal{O}(\xi^0),\mathcal{O}(\xi^2), \mathcal{O}(\xi^4)$ classes explicitely we have
\begin{align}
\mathcal{F}_{av}=&1
-\langle a_{1}^{2}\rangle
-\left[\langle a_{2}^{2}\rangle+2\langle \avec_{1}\avec^{T}_{3}\rangle-\frac{\langle a_{1}^{4}\rangle}{3}\right]+\sum_{k=3}^{\infty}\mathcal{O}(\xi^{2k})
\end{align}

Immediately we see that there is a collection of terms with equal magnitude arising from \emph{different orders} of the Magnus expansion (e.g. $a_{2}^{2}$ vs $a_{1}^{4}$).  The individual terms in the series expansion of the fidelity rely on time-domain correlation and cross-correlation functions and convolution with a multidimensional control matrix capturing the effect of the control operations. 

 The fidelity is thus expressed explicitly in terms of noise correlations and the control matrix. For instance,
\begin{align}\label{Eq:firstord}
&\langle a_{1}^{2}\rangle\nonumber\\
&=\sum_{i,j=x,y,z}\int^{\tau}_{0}dt_{2}\int^{\tau}_{0}dt_{1}
\langle\beta_{i}(t_{1})\beta_{j}(t_{2})\rangle \boldsymbol{R}_{i}(t_{1}) \boldsymbol{R}^{T}_{j}(t_{2})\nonumber\\
&=\sum_{i,j,k=x,y,z}\int^{\tau}_{0}dt_{2}\int^{\tau}_{0}dt_{1}
\langle\beta_{i}(t_{1})\beta_{j}(t_{2})\rangle R_{ik}(t_{1})R_{jk}(t_{2})
\end{align}
contains all two-point noise cross-correlation functions $\langle\beta_{i}(t_{1})\beta_{j}(t_{2})\rangle$, for $i,j\in\{x,y,z\}$, Higher-order terms contain multipoint correlation functions (this is determined by the sum of subscript indices, as they indicate the expansion-order of the error vector).

We rewrite these terms in the frequency domain, defining the Fourier transform $\mathcal{S}_{i_{1}...i_{n}}(\omega_{1},...,\omega_{n})$ of an $n$-point cross-correlation function via
\begin{widetext}
\begin{align}
\langle\beta_{i_{1}}(t_{1})\beta_{i_{2}}(t_{2})...\beta_{i_{n}}(t_{n})\rangle
\equiv\frac{1}{(2\pi)^{n}}\int d\omega_{1}...\int d\omega_{n} \mathcal{S}_{i_{1}...i_{n}}(\omega_{1},...,\omega_{n})e^{i(\omega_{1}t_{1}+...+\omega_{n}t_{n})}
\end{align}
The fidelity above can then be rewritten as
\begin{equation}\label{Eq:fidexpfreq}
\mathcal{F}_{av}=1-\sum_{n=2}^{\infty}
\left\{\frac{1}{(2\pi)^{n}}\sum_{i_{1}...i_{n}}\int d\omega_{1}...\int d\omega_{n}
\mathcal{S}_{i_{1}...i_{n}}(\omega_{1},...,\omega_{n})\mathcal{R}{i_{1}...i_{n}}(\omega_{1},...,\omega_{n})
\right\}
\end{equation}
where $\mathcal{R}_{i_{1}...i_{n}}(\omega_{1},...,\omega_{n})$ is determined solely by the control matrix and increases in complexity at higher order.  Explicit expressions for terms to arbitrary order are found in~\cite{GreenNJP2013}.
\end{widetext}


\subsubsection{First-order fidelity approximation}\label{SubSec:FirstOrderInfidelity}
Here we briefly explain the choice of fidelity metric used in the figures of the main text to produce the theory curves against which our experimental data is compared. Experimental fidelities are determined by measuring the brightness of the ion cloud after completing the control sequence, effectively yielding a projective measurement onto the $\ket{\uparrow}$ state. We denote this metric by $P_{\uparrow}(\tau) \in [\frac{1}{2},1]$ and refer to it as the \emph{state fidelity}, with lower and upper bounds corresponding to complete decoherence and perfect fidelity respectively. 

If the noise is sufficiently weak ($\xi^2\ll1$) we may truncate the series expansion for fidelity after the $\mathcal{O}(\xi^2)$ term yielding the  approximation
\begin{align}
\mathcal{F}_{\mathcal{O}(1)}=1
-\langle a_{1}^{2}\rangle.
\end{align}
\noindent Here the term which dominates the measured infidelity is $\langle\normalfont{a}^2_1\rangle:=\langle\EV_1(\tau)\EV_1^T(\tau)\rangle$, defined as the ensemble averaged modulus square of the first order error vector $\EV_1(\tau)$. Assuming wide sense stationarity, independence and zero mean of both noise fields $\Bd(t)$ and $\Ba(t)$ we may derive a spectral representation of $\langle\normalfont{a}^2_1\rangle$ of the form 
\begin{align}\label{FirstOrderInfidelityComputational}
&\langle a_1^2\rangle =\nonumber\\
&\frac{1}{2\pi}
\int_{-\infty}^{\infty}
\frac{d\omega}{\omega^2}
\Sd(\omega)
\Fd(\omega)+
\frac{1}{2\pi}
\int_{-\infty}^{\infty}
\frac{d\omega'}{\omega'^2}
\Sa(\omega') 
\Fa(\omega').
\end{align}
\noindent Here $\Sd(\omega)$ and $\Sa(\omega)$ denote the dephasing and amplitude noise PSDs. The dephasing $\Fd(\omega)$ and amplitude $\Fa(\omega)$ filter functions, on the other hand, capture the spectral response of the control sequence and are completely defined as functions of the control sequence. 

As the integrated noise content increases, however, higher-order error contributions must be included; neglecting to do leads to the unphysical result that $\mathcal{F}_{\mathcal{O}(1)}\le0$ when $\xi^2\ge1$. Although computation of all higher-order contributions is challenging we may gain some insight into the full expansion by considering terms of the form $\langle a_1^{2m}\rangle\equiv\langle a_1^2\rangle^m$ in each class $\mathcal{O}(\xi^{2m})$. This collection of terms is obtained by setting $a^2\rightarrow a_1^2$ in Eq. \ref{TraceFidelityCosExpansion}, effectively including only the first-order \emph{Magnus} expansion term in the expansion for \emph{Fidelity}, yielding 
\begin{align}
\label{Eq:FirstOrderFidelityExpansion1}\mathcal{F}'_{\mathcal{O}(1)}(\tau) &=\frac{1}{2}
\Big[
1+\sum_{m=0}^\infty(-1)^m\frac{2^{2m}}{(2m)!}\langle {a_1^2}\rangle^m
\Big]\\
\label{Eq:FirstOrderFidelityExpansion2}&=1
-\langle a_1^2\rangle
+\frac{\langle a_1^4\rangle}{3}
-\frac{2\langle a_1^6\rangle}{45}
+...
\end{align}

The oscillating sign of these terms is characteristic of the higher-order classes in converging to the true expresssion. To overcome the unphysicality of $\mathcal{F}_{\mathcal{O}(1)}$ as the noise content increases we employ a metric $\mathcal{F}_\chi$ with the physically reasonable properties that 
\begin{align}
&\mathcal{F}_{av}\approx \mathcal{F}_{\mathcal{O}(1)}\approx\mathcal{F}_{\chi},&\langle a_1^2 \rangle \ll 1\\
&\mathcal{F}_{\mathcal{O}(1)}\le \mathcal{F}'_{\mathcal{O}(1)} \le \mathcal{F}_{av},\mathcal{F}_\chi,&\langle a_1^2 \rangle \approx 1\\
&\mathcal{F}_{\chi} \rightarrow 1/2\rightarrow P_{\uparrow}(\tau),&\langle a_1^2 \rangle \gg 1
\end{align}

\noindent We may satisfy these conditions by noticing the qualitative resemblance between Eqs. \ref{Eq:FirstOrderFidelityExpansion1} and \ref{Eq:FirstOrderFidelityExpansion2} and the expansion for a simple exponential
\begin{align}
1
-\langle a_1^2\rangle
+\langle a_1^4\rangle
-\frac{2\langle a_1^6\rangle}{3}
+...&= \frac{1}{2}\left[1+
\sum_{m=0}^{\infty}(-1)^{m}\frac{2^m}{m!}\langle a_1^2\rangle^m
\right]\\
&=\frac{1}{2}\left[1+
\sum_{k=0}^{\infty}\frac{(-\chi(\tau))^m}{m!}
\right]
\end{align}
where we have defined $\chi(\tau)\equiv2\langle a_1^2\rangle$. Hence we use the following metric in calculating fidelities to be compared with experimental data
\begin{align}
\mathcal{F}_\chi = \frac{1}{2}\Big\{1+\exp[-\chi(\tau)]\Big\}
\end{align}

This approximation represents the \emph{first-order fidelity approximation}: it ignores higher-order cross correlations in the noise arising from higher-order Magnus contributions to the error vector, with diminishing overall magnitude (as given by the smallness parameter), but incorporates an approximation to higher-order terms important as the total noise-induced infidelity grows.  We work in this limit throughout this manuscript.


\subsubsection{Breakdown of the first-order fidelity approximation}

As described above, the first-order fidelity ignores higher-order terms expressed as nested-integrals over cross-correlations between noise along different directions, assuming weak noise.  As these contributions to gate infidelity grow in importance (for instance with $\alpha$) we expect the filter-transfer-function fidelity calculations to underestimate measured error in cases where the control has  filtered the noise to leading order.  

\begin{figure}[tp]
\includegraphics[width=7.5cm]{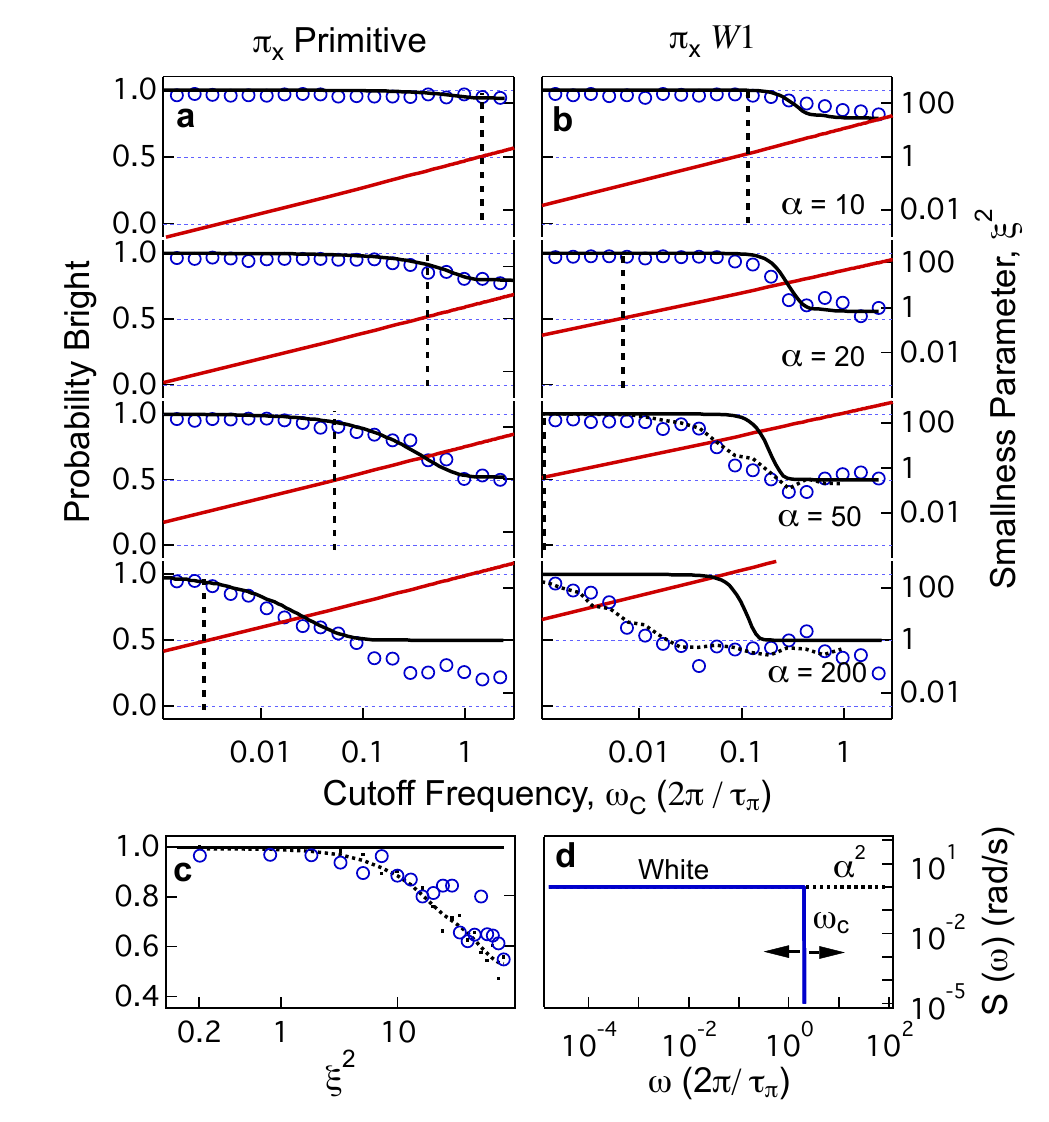}
\caption{\label{Fig:Pi_FFvalidation}Experimental validation of the breakdown of the first-order approximation to fidelity.  a)-b) Measurements of operational fidelity with engineered noise for rotation $\ket{0}\to\ket{1}$ as a function of dimensionless noise cutoff frequency.  Primitive rotation (a) and four-segment WAMF modulated rotation, $W1$, (b). Decay to value 0.5 corresponds to full decoherence.  Each data point is the result of averaging over 50 different noise realizations.  Maximum Rabi rate is fixed; filter $W1$ is conducted over time $4\tau_{\pi}$. Black dotted lines indicate the (smoothed) results of numerical integration of the Schroedinger equation, indicating that divergence between data and filter-function calculation is not due to experimental artefacts. Red lines (right axis) give $\xi^{2}$ employed in making the first-order filter-transfer function approximation. Vertical dashed lines indicate values of $\omega_{c}$ beyond which $\xi^{2}\geq 1$.  c) Measurements of $W1$ fidelity revealing a growing breakdown in agreement between filter-transfer function fidelity prediction and experimental data beyond $\xi^{2}\gtrsim1$ , taken for $\omega_{c}/2\pi=1.7\% \tau_{\pi}^{-1}$.  d) Schematic representation of the white noise power spectrum employed.}
\end{figure}

We measure the probability that a $\pi_{x}$-pulse drives qubit population from the dark state to the bright state, $\ket{0}\to\ket{1}$, as a function of the high-frequency cutoff, $\omega_{c}$, of a white dephasing bath (Fig.~\ref{Fig:Pi_FFvalidation}b).  As the high-frequency cutoff of the noise is increased and fluctuations fast compared to the control are added to the noise power spectrum, $S_{z}(\omega)$, errors accumulate reducing the measured fidelity.  The value of $\omega_{c}$ at which the fidelity drops from near unity decreases as a function of the noise strength, parametrized by $\alpha$.  In all cases for the primitive $\pi$ pulse the fidelity calculated using the filter transfer function matches the measured data well with \emph{no free parameters}.  

Performance is notably different when studying the four-segment WAMF $\pi$-pulse, $W1$, indicated in Fig.~\ref{Fig:FilterConstruction}d.  
 The WAMF construction provides \emph{first-order} filtering of time-dependent noise (red line in Fig.~\ref{Fig:FilterConstruction}b) (effectively cancelling terms proportional to $\langle a^{2}_{1}\rangle$), but does not provide suppression of higher-order terms in the Magnus expansion for fidelity which grow in importance with noise strength.  Unlike data for the primitive gate, as the noise strength increases  we observe a growing divergence between the measured fidelity and the fidelity calculated using the filter-transfer functions introduced above assuming a first order approximation (Fig.~\ref{Fig:Pi_FFvalidation}b).  
  
This phenomenon is not a function of total error magnitude, but instead occurs for $\xi^{2}\geq 1$ (red lines, right axis), a proxy measure indicating that we are not formally able to truncate the series expansion for fidelity at first order and must consider higher-order error contributions~\cite{GreenNJP2013}, including Magnus terms above $\langle a^{2}_{1}\rangle$.   These measurements therefore reveal the efficacy of noise filtering and quantitatively demonstrate the bounds of the first-order fidelity approximation as breakdown routinely occurs near the predicted value $\xi^{2}\geq 1$.  Notably, while formal convergence of this series requires $\xi^{2}\ll 1$, we find reasonable agreement between experiment and theory up to $\xi^{2}\sim 5$ (Fig.~\ref{Fig:Pi_FFvalidation}c). 

\subsection{Time-domain filter order vs. Magnus order}

We may formally indicate the functional dependence of the filter function on the control sequence by writing $F(\tau\omega) =F(\tau\omega;\CPSMatElem_n)$. Noise filtering (and hence error suppression) corresponds to minimizing the area under the filter transfer function in the spectral region where the noise PSD is non-negligible.  We therefore define a cost function over a user-defined frequency band which may take the form
\begin{align}\label{ErrorSuppressionFunctional}
A(\CPSMatElem_n)&:=\int_{\omega_{L}}^{\omega_c}d\omega F(\tau\omega;\CPSMatElem_n)
\end{align}
to diagnose the filtering effectiveness achieved by the control sequence $\CPSMatElem_n$; the smaller the integral $A(\CPSMatElem_n)$, the more effective the noise filtering in this band.  Having defined control sequences as continuous elements of the corresponding control space, for a given $n$ we may in principle construct a variational procedure on $\CtrlSpace_n$ to derive ``values'' of $\CPSMatElem_n$ satisfying a given cost function. 

The filter transfer function may be approximated by a polynomial expression $F(\omega\tau)\propto (\omega\tau)^{2p}$ for some $p$ near $\omega\approx0$. As $p$ increases the integral in \ref{ErrorSuppressionFunctional}, and hence the infidelity, decreases: the noise in the time domain is then said to be filtered to order $p-1$. 

Equivalently stated, a control sequence $\CPSMatElem_n\in\CtrlSpace_n$ filters time-dependent noise to order $p-1$ if $\CPSMatElem_n$ is a \emph{concurrent zero} of the first $p-1$ coefficients in the Taylor expansion\footnote{This procedure is valid for frequencies sufficiently lower than $1/\tau$ (the inverse of the total sequence duration).} of the filter transfer function about $\omega = 0$. 
\begin{align}\label{FFTaylorExpansionGeneral}
F(\omega\tau;\CPSMatElem_n) = \sum_{k=1}^{\infty}C_{2k}(\CPSMatElem_n)(\omega\tau)^{2k}.
\end{align}
The dependence of the expansion coefficients on our control parameters $\CPSMatElem_n$ has been made explicit, and we include only even powers of $\omega\tau$ due to the evenness of the filter transfer function. In this case $A(\CPSMatElem_n)\approx C_{2p}(\CPSMatElem_n)\frac{(\tau\omega_c)^{2p+1}}{2p+1}$ and the condition that 
\begin{align}\label{SuppressionConditions}
\frac{A(\CPSMatElem_n)}{C_{2p}(\CPSMatElem_n)}=\mathcal{O}\Big(\frac{(\tau\omega_c)^{2p+1}}{2p+1}\Big)
\end{align}
therefore implies the control sequence $\CPSMatElem_n$ filters noise to order $p-1$.  This effect is visualized through the slope of the filter transfer function in the stopband on a log-log plot (Fig.~\ref{Fig:FilterConstruction}b).  A high-order filter has a higher slope in this region, indicating improved suppression of time-dependent noise.

General filter design focuses on a band of interest, permitting spectral response to diverge outside of the spectral region of interest - for instance electrical filters in the microwave may appear transparent in the THz or Hz.  Therefore, in addition to the asymptotic, zero-frequency filter order $(p-1)$, we introduce a more general  metric capable of describing filter performance \emph{over an arbitrary spectral band}. The \emph{local filter order} $(p^*-1)$ establishes that the filter-transfer function is well approximated by $F_i\propto(\omega\tau)^{2p^*}$ over the band $[\omega_L,\omega_c]$.  It is this more narrowly defined metric that is used in most practical filter-design tasks, including those undertaken above.

The performance of filter-order ($p-1$) or local filter order ($p^{*}-1$) for time-dependent noise described above must be distinguished from the order of error suppression for quasistatic errors in the Magnus expansion.  The latter measure is typically used in NMR literature to the pulse sequences designed to compensate for quasistatic errors. In this regime the time dependence of the dephasing (amplitude) noise fields reduces to constants $\Bd\; (\Ba)$ and the Magnus expansion terms $\Phi^\text{(DC)}_\mu$ are evaluated strictly as time integrals over \emph{ideal} control operations scaled by powers of the offset magnitude $\Bd^\mu\;(\Ba^\mu)$. A pulse sequence for which $\Phi^\text{(DC)}_1 = ...= \Phi^\text{(DC)}_{\mu-1} = 0$ is then said to compensate offset errors to order $\mu-1$. In this case the total error operator satisfies $\Phi^\text{(DC)}(\tau)=\mathcal{O}(\Phi^\text{(DC)}_\mu)$ and is dominated by the residual error proportional to the $\mu$th power in the offset magnitude.  

\emph{High-order error suppression in the Magnus expansion does not imply high-order time-domain noise filtering}. Table \ref{Table:OESMetrics} reveals the importance of not conflating these two measures when assessing the performance of a control sequence against static vs stochastic errors.  The upper panel compares the two performance measures for some well-known phase-modulated NMR sequences, the naming conventions for which are consistent with the review by Merrill and Brown\cite{MerrillArXv2012}. For completeness, in the lower panel we also make the comparison for the novel control sequences derived in this paper. 
\begin{table}[tp]
  \centering
\begin{tabular}{c|cc||cc|}
\cline{2-5}
 & \multicolumn{2}{c||}{Amplitude Errors}  &\multicolumn{2}{c|}{Dephasing Errors}\\
\cline{2-5}
 &$\mu-1$ &  $p-1$  &$\mu-1$ &  $p-1$  \\
\hline
\multicolumn{1}{|c|}{SK1}& $1$ &$1$  & $0$ &$0$   \\
\multicolumn{1}{|c|}{P2}& $2$ &$1$  & $0$ &$0$   \\
\multicolumn{1}{|c|}{B2}& $2$ &$1$  & $0$ &$0$   \\
\multicolumn{1}{|c|}{C1}& $0$ &$0$  & $1$ &$1$   \\
\multicolumn{1}{|c|}{C2 ($\pi$)}& $0$ &$0$  & $2$ &$0$   \\
\hline\hline
\multicolumn{1}{|c|}{W1}& $0$ &$0$  & $1$ &$1$   \\
\multicolumn{1}{|c|}{W2}& $0$ &$0$  & $1$ &$2$   \\
\multicolumn{1}{|c|}{UWMG$_{1,SK1}$}& $1$ &$1$  & $1$ &$1$   \\
\hline
\end{tabular}
\caption{Comparision between suppressing \emph{stochastic} errors to order $p-1$ (filter order) and compensating for \emph{static offset} errors to order $\mu-1$ (Magnus order). Naming conventions for NMR sequences in top panel are consistent with the review article by Merrill and Brown \cite{MerrillArXv2012}.  The final entry corresponds to a concatenated construction described below.\label{Table:OESMetrics}}
\end{table}

Later we will return to the question of time-domain filter order and introduce a set of analytic design rules for filter construction based on the characteristics of our selected basis functions - the Walsh functions.


\subsection{Walsh basis functions}

We impose physically motivated constraints on the form of $\CPSMatElem_n$ in order to reduce the search to a manageable subspace of $\CtrlSpace_n$, and elect to synthesize control sequences from the Walsh basis functions. The set of Walsh functions $\Walsh_k:[0,1]\rightarrow\{\pm1\}$, $k\in\mathbb{N}$ form an orthonormal-complete family of binary-valued square waves defined on the unit interval and are the \emph{digital analogues} of the sines and cosines in Fourier analysis.  Since their formulation in the first half of the twentieth century, Walsh functions have played an important role in scientific and engineering applications. Their development and utilzation has been strongly influenced by parallel developments in digital electronics and computer science since the 1960s, with Walsh-type transforms replacing Fourier transforms in a range of engineering applications such as communication, signal processing, image processing, pattern recognition, noise filtering and so forth\cite{Walsh_Beauchamp}. 


We summarize the relevant mathematical details of the Walsh basis, outlining two equivalent representations, \emph{Paley ordering} and the \emph{Hadamard representation}, both of which are useful to understand the Walsh control space. 


\subsubsection{Paley ordering}\label{SubSec:PaleyOrdering}

\begin{figure}
\centering
\includegraphics[width=0.9\columnwidth]{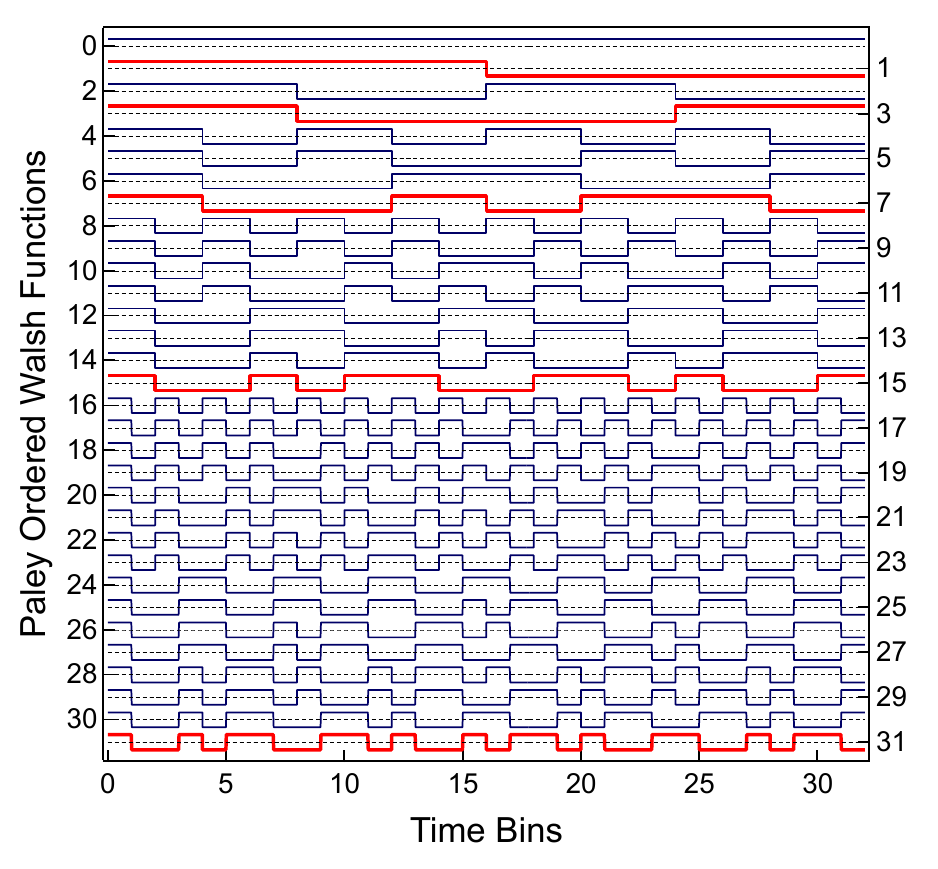}
\caption{The first 32 Walsh functions in Paley ordering.}  \label{Fig:WalshFunctions}
\end{figure}

The Walsh functions are aperiodic and hence do not admit to the unique ordering according to increasing \emph{frequency} characteristic of the sinusoids in the Fourier basis. A number of differnt orderings exist with associated definitions of the basis elements. We employ the \emph{Paley ordering} in which basis functions are generated from products of \emph{Rademacher functions}, a family of square waves defined by 
\begin{align}\label{Rademacher}
R_j(x) := \text{sgn}\big[\sin(2^j\pi x)\big],\hspace{0.8cm} x\in[0,1],\hspace{0.8cm} j\ge0,
\end{align}
switching between $\pm1$ at rate $2^j$. The Walsh function of Paley order $k$,  denoted $\PAL_k(x)$, is then defined by
\begin{align}\label{PALgeneration}
\PAL_k(x) = \prod_{j=1}^{m} R_j(x)^{b_j}
\end{align}
where $(b_m,b_{m-1},...,b_1)_2$ is the binary representation of $k$. That is, $k = b_m2^{m-1}+b_{m-1}2^{m-2}+...+b_{1}2^0$, 
where  $b_m\equiv1$ is defines the most significant binary digit. Hence $R_j(x)$ is a factor of $\PAL_k(x)$ whenever $b_j$ is a nonzero  binary digits of $k$. The total number of nonzero $b_j$'s in $k$ define the \emph{Hamming wieght}, denoted by $r$. We write $m(k)$ and $r(k)$ when it is desirable to emphasize that both $m$ and $r$ are functions of $k$. For illustration, the first 32 Walsh functions in the Paley ordering are shown in Fig. \ref{Fig:WalshFunctions}.
%


\subsubsection{Hadamard representation}\label{SubSec:HadamardRepresentation}

For our purposes we require an expression for the piecewise-constant structure of an arbitrary superposition of Walsh functions, and therefore desire \emph{a priori} knowledge of the locations of their various zero crossings. A general expression, however, is difficult due to the aperiodicity of the Walsh functions. It is convenient instead to use the Hadamard representation in which any continuously defined basis member $\PAL_k(x)$ projects completely onto a digital vector in $\mathbb{R}^{2^n}$ provided $m(k)\le n$, which is true for the $2^n$ Paley orders in the set $k\in\{0,1,...,2^n-1\}$. Since these vectors have dimension $2^n$ and inherit the orthogonality of the $\PAL_k(x)$ they therefore form a \emph{discrete} Walsh basis spanning $\mathbb{R}^{2^n}$. Such a projection is clearly possible since the fastest modulation rate in $\PAL_k(x)$ derives from the periodicity of $R_{m(k)}(x)$, which switches sign $2^{m(k)}$ times over $x\in[0,1]$. The projection then involves partitioning the domain into $2^n$ bins and asociating the value of $\PAL_k(x)$ in the $j$th bin to the $j$th element $P^{(k)}_j\in\{\pm1\}$ of the discrete digital vector 
\begin{align}\label{VectorizedPAL}
\boldsymbol{P}^{(k)}_{2^n} = \left[\begin{array}{cccc}P^{(k)}_1,&\hspace{0.2cm}P^{(k)}_2,&... &P^{(k)}_{2^n}\end{array}\right].
\end{align}

Using the so-called \emph{Sylvester construction}~\cite{Horadam2007}, the $2^n$-dimensional Hadamard matrix $H_{2^n}$ is generated recursively by 
\begin{align}
&H_{2^n} = \left[\begin{array}{cc}H_{2^{n-1}} & H_{2^{n-1}}\\H_{2^{n-1}}&-H_{2^{n-1}}\end{array}\right] = S^{\otimes n}\\
&S = \left[\begin{array}{cc}1 & 1\\1 &-1\end{array}\right],\hspace{0.5cm} H_1 = 1
\end{align} 
where $S$ is the Sylvester matrix, and $\otimes n$ denotes $n\ge1$ applications of the Kronecker product. In this construction $\boldsymbol{P}^{(k)}_{2^n}$ defines the $i(k)= 1+\sum_{j=1}^{m(k)}b_j2^{n-j}$  column (row) of $H_{2^n}$. The orthogonality of the Walsh basis is thereby reflected in the property that $H_{2^n}H_{2^n}^T = 2^nI$, implying the orthogonality of the Hadamard matrices. 

This representation is particularly useful for efficiently constructing Walsh-synthesized waveforms. Consider an arbitrary function $f(x) = \sum_{k = 0}^N \ampP_k\PAL_k(x)$ synthesized in the Walsh basis where $N$ sets the highest (Paley) ordered function in the construction. Then, from the above discussion, all Walsh functions in this synthesis projected onto a Hadamard matrix of dimension $\ge 2^{m(N)}$, with $\MinDim = 2^{m(N)}$ giving the minimal sufficient dimension. A discrete representation of the function $f(x)$ therefore exists as a projection onto the column space of $H_\MinDim$ by writing
\begin{align}\label{fHadRep}
&\boldsymbol{f} = H_{_\MinDim}\boldsymbol{\ampH}.
\end{align}
The column vector $\boldsymbol{\ampH}=\left[ \begin{array}{cccc}  \ampH_1, & \ampH_2,&...&\ampH_{_\MinDim}\end{array} \right]^T$ contains the reordered Paley amplitudes $\ampP_k$ reoardered under the change of basis map $i(k)$ specified b   
\begin{align}
\label{HadAmpVecComponent}\ampH_{i(k)} &=\begin{cases}\ampP_k&\text{for}\hspace{0.5cm}0\le k\le N\\
0&\text{for}\hspace{0.5cm}N<k< \MinDim\end{cases}.
\end{align}
The vector $\boldsymbol{f}=\left[ \begin{array}{cccc}  f_1, & f_2,&...&f_{_\MinDim}\end{array} \right]^T$ so generated then represents the piecewise constant structure of $f(x)$, with $f_j$ giving the value taken by $f(x)$ on the $j$th of $\MinDim$ equal subintervals partitioning $x\in[0,1]$.


\subsection{Walsh synthesis for amplitude modulated filters}\label{Sec:WAMGs}

Any square integrable function $f(x)$ on the interval $[0,1]$ has a unique \emph{spectral decomposition} in the \emph{Walsh basis} 
\begin{align*}
&f(x) = \sum_{k=0}^\infty \ampP_k\Walsh_k(x)\hspace{0.25cm}\iff\hspace{0.25cm}\ampP_k := \int_0^1f(x)\Walsh_k(x)dx.
\end{align*}

\noindent We consider a control regime referred to as \emph{single-axis amplitude-modulation} defined by 
\begin{align}
\label{AmplitudeModulationDefinition}&\CPSMatElem_n\cong\{(\tau_l,\theta_l)\}_{l=1}^n, \hspace{0.25cm}\phi_l = \phi_0\hspace{0.25cm}\forall l\in\{1,...,n\}
\end{align}
where the waveform structuring the total angle swept out in each pulse segment, $\theta_{l}$, is based on a linear superposition of well-defined square waves known as Walsh functions. We refer to these sequences as \emph{Walsh amplitude modulated filters} (WAMFs), and speak of searching over the \emph{Walsh-modulated control subspace}. 

Using the above framework, we may efficiently construct arbitrary WAMFs for arbitrary pulse-segment envelopes.  We define a time-varying Rabi rate $\Omega_l(t)$, $l\in\{1,...,\MinDim\}$ over the time period $t\in[t_{l-1},t_l]$ of each segment, $l\in\{1,...,\MinDim\}$, $t\in[t_{l-1},t_l]$.  Each segment has duration $\tau_l=\tau/\MinDim$ and generates a  \emph{total rotation angle} for each segment, $\theta_l$ given by the integral 

\begin{align}
\theta_{l}=\int_{t_{l-1}}^{t_l}\Omega_{l}(t)dt
\end{align}

\noindent We now treat the rotation angles $\theta_{l}$ as parameters by which to optimize filter performance.  For efficient filter construction, however, it is convenient to instead transform this optimization over $\theta_{l}$ to an optimization over a Walsh spectrum.  This is achieved by writing$\theta_l =  \theta_l(\WAMampP_0, \WAMampP_1,...,\WAMampP_N)$ with the dependence on the Walsh spectra defined by the Hadamard-matrix equation  
\begin{align}\label{Eq:ThetaSynthesis}
\vec{\boldsymbol{\theta}} = \big(\theta_1,\hspace{0.1cm},\theta_2,...,\theta_\MinDim\big)^T = (\tau/\MinDim)H_\MinDim\boldsymbol{\WAMampH}.
\end{align}
Defined in this way, the $\MinDim$-segment arbitrary-envelope construction achieves total gate-rotation angle $\Theta = \sum_{l=1}^n\theta_l = \WAMampP_0\tau$, completely determined by the spectral amplitude of $\PAL_0$. All symmetry-based design rules carry over, regardless of the modulation envelope for an individual pulse segment.

\subsubsection{Square pulses}
The special case of square pulse segments is treated here as it allows a reduction in synthesis complexity and is compatible with many experimental systems.  We may replace the time-dependent Rabi rate $\Omega(t)$ over a single segment with a piecewise-constant (over a single-segment) construction used in Walsh synthesis over a complete pulse sequence
\begin{align}\label{WalshAmplitudeSynthesis}
\Omega(t) = \sum_{k=0}^N\ampP_k\PAL_k(t/\tau),\hspace{1cm} t\in[0,\tau].
\end{align}
This permits synthesis over the Rabi rate per segment rather than the total rotation angle, which is often simpler in experimental settings.  Substituting $\Omega(t/\tau)$ for $f(x)$ in Eq. \ref{fHadRep} we obtain $\mathbf{\Omega} = H_{_\MinDim}\boldsymbol{\ampH}$. The vector $\mathbf{\Omega}=\left[ \begin{array}{cccc}  \Omega_1, & \Omega_2,&...&\Omega_{_\MinDim}\end{array} \right]^T$ thus defines a sequence of modulated Rabi rates each functionally dependent on the Walsh amplitudes
\begin{align}
\Omega_l = \Omega_l(\ampP_0, \ampP_1,...,\ampP_N),\hspace{1cm}l\in\{1,...,\MinDim\}. 
\end{align}
The WAMF is then defined by explicitly writing $\boldsymbol{\Omega}$ as an additional column leading the representation of Eq.~\ref{TemplateCPSMatrix}, yielding the form 
\begin{align}\label{WAGMTemplateCPSMatrix}
\CPSMatElem_\MinDim\hspace{0.25cm}=\hspace{0.25cm}
\begingroup
  \renewcommand*{\arraystretch}{1.5}%
  \kbordermatrix{
           & \Omega_l           & \theta_l & \tau_l                     & \phi_l     \cr
    P_1    & \Omega_1             & \frac{\Omega_1\tau}{\MinDim}& \frac{\tau}{\MinDim}               & \phi_0 \cr
    P_2    & \Omega_2             & \frac{\Omega_2\tau}{\MinDim}&\frac{\tau}{\MinDim}              & \phi_0 \cr
    \vdots & \vdots        & \vdots        & \vdots        &  \vdots  \cr
    P_\MinDim    & \Omega_\MinDim            & \frac{\Omega_\MinDim\tau}{\MinDim}  & \frac{\tau}{\MinDim}               & \phi_0 \cr
  }%
\endgroup
\end{align}

Here the degree of freedom associated with $\tau_l$ has apparently been removed. This reflects the fact that the choice of $\tau_l$ has been transformed into the choice of Walsh basis functions in the synthesis, each contributing its characteristic temporal profile. The remaining degrees of freedom reside in functional dependence of $\Omega_l$ on the Walsh spectrum and our variational search is thus limited to the subspace of $\CtrlSpace_\MinDim$ effectively spanned by $\boldsymbol{\ampP}$. 

Negative Walsh spectral amplitudes may produce negative valued Rabi rates under a linear superposition. For a pulse of the form $P_l = \exp\big[-i\Omega_l\tau_l\sigphiL/2\big]$ the negative sign of $\Omega_l$ may be absorbed into the spin operator $\phi_l$ corresponds physically to the application of a $\pi$ phase shift int he diriving field. This follows from the fact that $\hat{\sigma}_{\phi+\pi} = -\hat{\sigma}_\phi$, which is clear from the definition of our spin operator in Eq. \ref{SpinOperatorDefinition}. Thus including negative-valued Walsh spectral amplitudes generally produces single axis control only up to a sign change.

\subsubsection{Gaussian pulse envelopes}
The square-envelope pulses studied experimentally in the main text are easy to generate in our experimental system but may prove difficult in other settings where abrupt amplitude shifts at timestep-edges produce significant pulse distortion.  Here we show that achieving Walsh synthesized filters using a common Gaussian pulse envelope yields comparable results with a simple re-optimization of Walsh-synthesis coefficients.

The square amplitude-modulated waveform is here replaced with a smoothly varying pulse envelope in each segment, each associated with a specific rotation angle $\theta_l$ subject to optimization.  We assume a Gaussian profile $G_l(t;\mu_l,\sigma_l)$ defined on $t\in[t_{l-1},t_l]$ with mean $\mu_l$ and standard deviation $\sigma_l$. Specifically, we construct

%
\begin{align}
G_l(t;\mu_l,\sigma_l)&=\frac{\theta_l}{C_l\sigma_l\sqrt{2\pi}}\exp\Big[{-\frac{(t-\mu_l)^2}{2\sigma_l^2}}\Big]\\
\mu_l &= \frac{t_{l-1}+t_l}{2}\\
\sigma_l &=  g\tau/\MinDim
\end{align}
with $\mu_l$ the segment midpoint and $\sigma_l$ expressed as a multiple $g$ of the segment duration. The normalizing factor
\begin{align}
C_l:=\int_{t_{l-1}}^{t_l}\frac{1}{\sigma_l\sqrt{2\pi}}\exp\Big[{-\frac{(t-\mu_l)^2}{2\sigma_l^2}}\Big]dt
\end{align}
is included to ensure the total rotation implemented by the Gaussian pulse in the $l$th segment is given by $\int_{t_l}^{t_{l-1}} G_l(t;\mu_l,\sigma_l)dt = \theta_l$. We now impose the same structure on the segment rotations $\theta_l$ as presented above in Eq.~\ref{Eq:ThetaSynthesis}.  Defined in this way, the $\MinDim$-segment Gaussian-pulse sequence shares with the square WAMF construction the property that the total gate rotation angle $\Theta = \sum_{l=1}^n\theta_l = \WAMampP_0\tau$ is completely determined by the spectral amplitude of $\PAL_0$. 

We may therefore construct a Gaussian-pulse variation on any candidate WAMF such that, having set $g$ to some value relevant to the control hardware, the smooth pulse sequence remains strictly parametrized in the Walsh spectrum $\boldsymbol{\WAMampP}$. In particular, filter optimization proceeds in the same manner as for ordinary Walsh-modulated control by minimizing the cost function with respect to the Walsh spectrum.


\subsection{Analytic design rules}
An advantage of Walsh synthesis is that the well-defined spectral properties and symmetries of the Walsh functions may be employed to further restrict the search space available for filter construction.

First, in practice the achievable filter order \emph{over the entire stopband} is limited by the number of constituent control operations; one may achieve higher $p$ at the cost of higher $n$. The maximum achievable value of $p$ for a given filter is set by the power-law expansion of the filter for the single Walsh function with the highest Paley order for a given $n$.  As has been shown previously, all Walsh functions with given Hamming weight of the Paley order have the same power-law expansion near zero frequency~\cite{HayesPRA2011}. Therefore, in principle, every doubling of $n$ increases the maximum achievable time-domain filter order by one.  The Walsh functions highlighted in red in Fig~\ref{Fig:FilterConstruction}c and Fig.~\ref{Fig:WalshFunctions} represent those with the highest Paley-order Hamming weight for a given $n$.  Nonetheless we find that in general we are able to construct filters with higher order than prescribed \emph{over narrow regions in the stopband}, as a result of Walsh synthesis (see the multiple slopes for the blue line in Fig~\ref{Fig:FilterConstruction}b).  


In filter construction we may further constrain the form of a candidate pulse sequence by imposing required physical properties on the sequence, such as fixing the total rotation angle of the Bloch vector in order to implement a target logic operation.  In order to proceed we then partition the Walsh spectrum $\boldsymbol{\ampP}\equiv(\boldsymbol{\ampP}_\nu,\boldsymbol{\ampP}_\rho)$ into spectral amplitude classes $\boldsymbol{\ampP}_\nu$ and  $\boldsymbol{\ampP}_\rho$ to be treated as \emph{variational} and \emph{fixed} parameters respectively. Fixed parameters set the physical state transformation of interest while the remaining unconstrained components in $\boldsymbol{\ampP}_\nu$ serve as tuning parameters by which to minimize $A(\boldsymbol{\ampP}_\nu;\boldsymbol{\ampP}_\rho)$.

The primary constraint in WAMF constructions is that the total rotation angle executed depends only on the value of $\ampP_0$, the zeroth order spectral component; it sets the effective average Rabi rate for the WAMF. This can be seen as follows. First observe all Walsh functions of higher than zeroth order are \emph{balanced} in the sense that $\int_0^1\PAL_k(x)dx = \delta_{0k}$. For the control field defined by Eq. \ref{WalshAmplitudeSynthesis} the total gate rotation angle $\Theta = \int_0^\tau \Omega(t) dt$ then takes the form
\begin{align*}
\Theta  &= \int_0^\tau\sum_{k=1}^N\ampP_k\PAL_k(t/\tau)dt\\
 &= \tau\sum_{k=1}^N\ampP_k\int_0^1\PAL_k(x)dx\\
& = \tau\sum_{k=1}^N\ampP_k\delta_{0k} = \ampP_0\tau.
\end{align*}
In this case the net gate rotation $\theta = \Theta\hspace{0.1cm}\text{mod}\hspace{0.1cm}2\pi$ is given by 
\begin{align}\label{NetBlochRotationConstraint}
\theta = \ampP_0\tau\hspace{0.1cm}\text{mod}\hspace{0.1cm}2\pi
\end{align} 
implying the necessary constraint on $\ampP_0$ in order to achieve a desired $\theta$.

Next, we observe that the Walsh functions have distinct parity, but that filter constructions mandate symmetric constructions in order to enact a target operation and provide effective noise cancellation.  The result is that odd-parity Walsh functions may generally be excluded from the variational search.  While this is not necessarily strictly required (multiple odd parity Walsh functions may in principle be added with opposite signs to produce net symmetric constructions), it is convenient and effective to restrict the synthesis space to the so-called $\text{CAL}$ subset of the Walsh functions.

Our reduced search problem may then be represented formally by replacing $\CPSMatElem_\MinDim\rightarrow(\boldsymbol{\ampP}_\nu
,\boldsymbol{\ampP}_\rho)$ in Eq. \ref{ErrorSuppressionFunctional} to obtain
\begin{align}\label{WAMGAreaFunctional}
A(\boldsymbol{\ampP}_\nu;\boldsymbol{\ampP}_\rho)&:=\int_0^{\omega_c}d\omega F(\tau\omega;\boldsymbol{\ampP}_\nu
,\boldsymbol{\ampP}_\rho)
\end{align}
with the variational search now restricted to the subspace spanned by $\boldsymbol{\ampP}_\nu$ with the $\boldsymbol{\ampP}_\rho$ held constant. 


\subsection{First order WAMFs}\label{SubSec:FirstOrderWAMGs}
As a first application of the above result, we derive a family of nontrivial gates decoupled to first order against dephasing noise by constructing a pulse sequence from the synthesis $\theta(t) = \frac{\tau}{4}\Big(\ampP_0\PAL_0(t/\tau)+\ampP_3\PAL_3(t/\tau)\Big)$. Note that $\theta(t)$ is only formally defined at the end of pulse segments.  That is, we set $\boldsymbol{\ampP}_\rho \equiv \ampP_0$ and $\boldsymbol{\ampP}_\nu  \equiv \ampP_3$. In this case $N = 3$ and $\MinDim = 4$, so the minimal pulse duration is $\tau/4$. Using Eq. \ref{HadAmpVecComponent} we obtain $\boldsymbol{\ampH}=\left[ \begin{array}{cccc}  \ampH_1, & \ampH_2,&\ampH_3,&\ampH_4\end{array} \right]^T = \left[ \begin{array}{cccc}  \ampP_0, & 0,&0,&\ampP_3\end{array} \right]^T$, yielding the minimal Hadamard representation
\begin{align}\label{WAMGO1HadamardRepresentation}
&\boldsymbol{\theta} = \frac{\tau}{4}\left[\begin{array}{cccc}1&1&1&1\\1&-1&1&-1\\1&1&-1&-1\\1&-1&-1&1\end{array}\right]\left[\begin{array}{cccc}  \ampP_0\\ 0\\ 0\\ \ampP_3\end{array} \right] = \frac{\tau}{4}\left[\begin{array}{cccc}  \ampP_0+\ampP_3\\ \ampP_0-\ampP_3\\ \ampP_0-\ampP_3\\\ampP_0+\ampP_3\end{array} \right]
\end{align}
\noindent These sequences therefore span the control subspace parametrized by $(\tau=1)$
\begin{align}\label{WAGMOrder1CPSMatrix}
\text{WAMF }\mathcal{O}(1)\hspace{0.25cm}=\hspace{0.25cm}
\begingroup
  \renewcommand*{\arraystretch}{1.5}%
  \kbordermatrix{
           & \Omega_l   & \theta_l           & \tau_l                   & \phi_l     \cr
    P_1    &\WAMGRabiPlus & \frac{\WAMGRabiPlus}{4}              &\frac{1}{4}             & 0 \cr
    P_2    & \WAMGRabiMin    & \frac{\WAMGRabiMin}{4}         &\frac{1}{4}             & 0 \cr
    P_3   &\WAMGRabiMin    & \frac{\WAMGRabiMin}{4}          & \frac{1}{4}              & 0 \cr
       P_4   &\WAMGRabiPlus    & \frac{\WAMGRabiPlus}{4}          & \frac{1}{4}              & 0 \cr
  }%
\endgroup
\end{align}
where we have defined $\ampP_{\pm} = \ampP_0\pm\ampP_3$.  This choice is motivated by the fact that Paley order $k = 3$ corresponds to the lowest order non-constant Walsh function with even symmetry about $\tau/2$ (Fig. \ref{Fig:WalshFunctions}). Hence \ref{WAMGO1HadamardRepresentation} is the simplest Walsh modulated form which includes the zeroth order Walsh function in the synthesis and which possesses time-reversal symmetry about the sequence midpoint. The former property ensures a nontrivial gate angle is executed. The latter is chosen due to the observation in dynamic decoupling literature that using time-symmetric building blocks often improves the performance of the sequence compared to sequences formed by time-asymmetric building blocks\cite{Lidar2013}\cite{Suter_Interleaved}.  In this construction we have maintained a leading column of rabi rates, $\Omega_{l}$, as would be appropriate for the square-pulse forms used in the main text. 

\subsubsection{Gaussian pulse construction}

 \begin{figure}[bp]
\includegraphics[width=7cm]{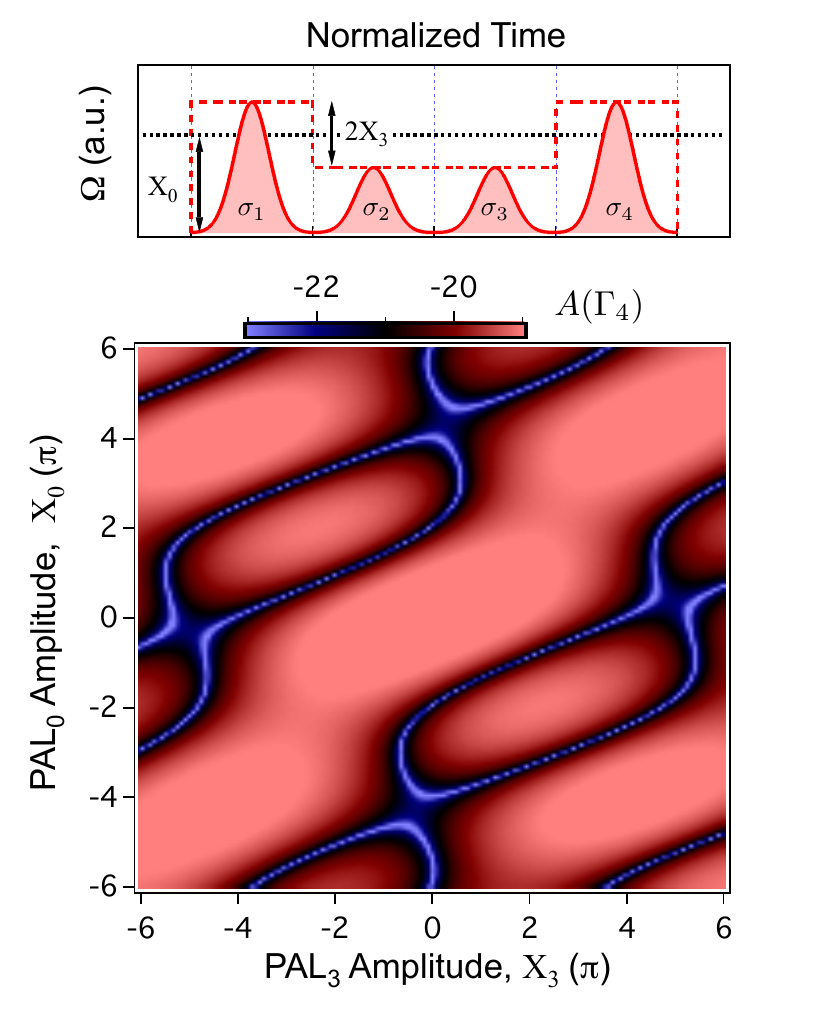}
\caption{\label{Fig:2DCostFunctionGaussianPulses}Construction of the first-order Walsh amplitude modulated dephasing-suppressing filter using Gaussian-shaped pulse segments.   a) Schematic representation of Walsh synthesis for a four segment gate of discrete Gaussian segments.  Walsh synthesis determines the overall amplitude of Gaussian pulses with fixed duration and standard deviation, setting the effective pulse area in each segment.     b) Two-dimensional representation of the integral metric defining our target cost function, $A(\CPSMatElem_{4})$ integrated over the stopband $\omega\in [10^{-9}, 10^{-6}]\tau^{-1}$. Areas in blue minimize $A(\CPSMatElem_{4})$, representing effective filter constructions. The $X_{0}$ determines the net rotation enacted in a gate while $X_{3}$ determines the modulation depth, as represented in a). }
\end{figure}

Adding to the results presented in the main text, constructing the 4-segment filter $W1$ using square pulse segments, we examine the Gaussian-pulse variation here.  The cost function $A_z(\WAMampP_3;\WAMampP_0) = \int_{\omega_L}^{\omega_c}d\omega \Fd(\omega\tau;\WAMampP_3;\WAMampP_0)$ may be computed by partitioning the time domain into a large number $N_s$ of subintervals on which the continuous Gaussian envelope is treated as approximately constant. Fig. \ref{Fig:2DCostFunctionGaussianPulses} shows a two-dimensional representation of $A_z(\WAMampP_3;\WAMampP_0)$ integrated over the interval $\omega\in [10^{-9}, 10^{-6}]\tau^{-1}$, with $g = 1/6$ and $N_s = 100$. The value of $\text{Log}_{10}\big[A_z(\WAMampP_3;\WAMampP_0)\big]$ is indicated by the color scale. Total sequence length is normalized to $\tau=1$ in this data, so the total gate rotation angle $\Theta \equiv \WAMampP_0$ is given directly by the $X_0$-axis. Regions in blue represent effective (first-order) filter constructions, where the cost function is minimized.  

We conclude useful filter construction using Gaussian pulses is a simple matter of re-optimization in the Walsh-synthesis framework. This is readily achieved using a Nelder-Mead optimization of $A_z(\WAMampP_3;\WAMampP_0)$ for any particular choice of $g,\omega_L,\omega_c,\WAMampP_0$ or $N_s$, in a manner precisely the same as for square envelopes.


\subsection{Universal Filters by Concatenation\label{Sec:Concatenation}}

Phase-modulated sequences robust against amplitude noise may also be found in the Walsh basis, yielding \emph{Walsh phase-modulated filters} (WPMFs) analogous to WAMFs, and implementing arbitrary target rotations $\theta$.   There are a variety of techniques to construct such WPMFs, but we use analytic design rules in which a target rotation is performed (with some error due to noise), and phase-modulated segments are added in order to produce the net filtering effect.  The simplest WPMF adds two segments phase modulated according to Walsh function $\PAL_{1}$ with coefficient $X_{1}$, subject to the constraint that one enacts the desired driven rotation by $\theta$.  

The result of this approach yields a WPMF that is identical to the NMR sequence SK1, with value $X_{1}=\cos^{-1}(-\theta/4\pi)\equiv\phiSK(\theta)$~\cite{MerrillArXv2012}, as represented

\begin{align}\label{SK1CPSMatrix}
\CPSMatElem_3^{(\text{SK1})}\hspace{0.1cm}&=\hspace{0.1cm}
\begingroup
  \renewcommand*{\arraystretch}{1.5}%
  \kbordermatrix{
           & \Omega_l           & \theta_l    & \tau_l                  & \phi_l     \cr
    P_1    &  \Omega_0            & \theta&\tau_\theta               & 0 \cr
    P_2    & \Omega_0             &2\pi &\tau_{2\pi}            & \phiSK \cr
    P_3    & \Omega_0             &2\pi   &\tau_{2\pi}             & -\phiSK \cr
  }%
\endgroup
\end{align}
\begin{align}
\Omega_0 = \frac{\theta+4\pi}{\tau},\hspace{0.5cm}\tau_\theta = \frac{\theta}{\Omega_0},\hspace{0.5cm}\phiSK(\theta): = \cos^{-1}\Big(-\frac{\theta}{4\pi}\Big).
\end{align}

\noindent Note that the Walsh timing construction only holds in the two correction steps represented above.  Following a similar route allows one to construct a sequence with modulation given by $\PAL_{3}$ which is formally identical to the three-segment (four-timestep) phase modulation given by gate $P2$~\cite{MerrillArXv2012}.  

\begin{figure}[tp]
\includegraphics[width=8.5cm]{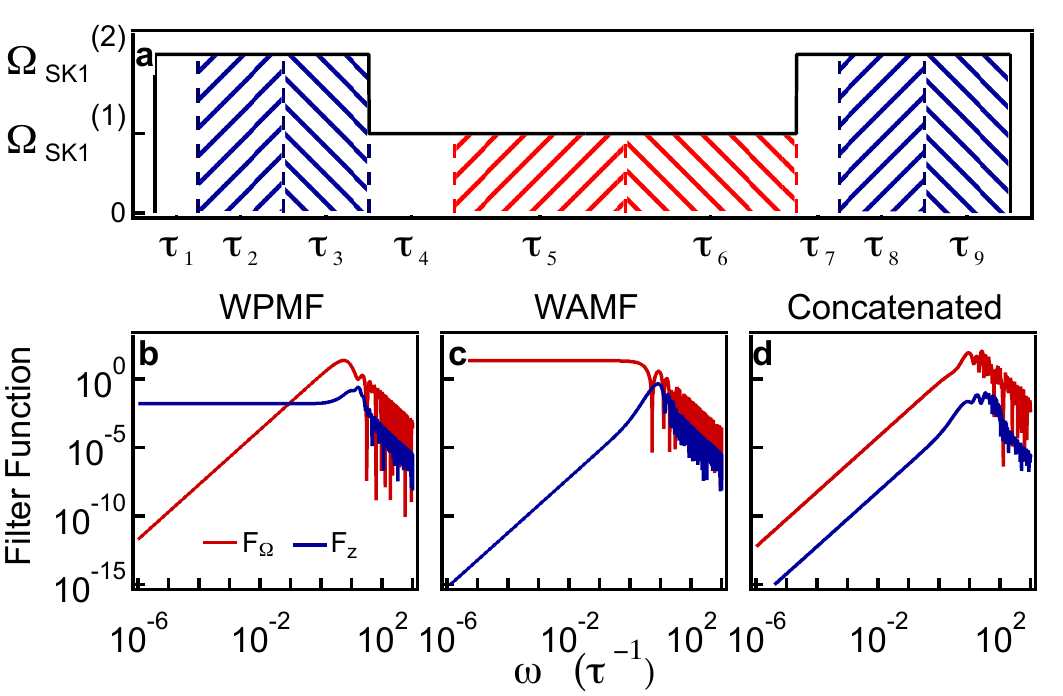}
\caption{\label{Fig:Concatenation}Concatenation scheme for universal noise suppression. a) Concatenation of WPMF$\equiv$SK1 within first order WAMF sequence yielding UWMF$_{1,SK1}$. White fill indicates rotations enacted with $\phi = 0$; orientation of hatching denotes SK1 phase flips $\phi = \pm \phiSK$. b) Filter functions for WPMF$\equiv$SK1 sequence. c) Filter functions for four-segment WAMF sequence, $W1$. d) Filter functions for concatenated sequence.}
\end{figure}

These WPMF sequences perform as first-order time-dependent noise filters, captured in the form of $\Fa(\tau\omega)$, and noted in Table ~\ref{Table:OESMetrics} (column 2).   For example, filter functions for the WPMF that is equivalent to SK1 are shown in Fig. ~\ref{Fig:Concatenation}b, revealing first-order filtering of amplitude noise, but not dephasing noise.


We may now concatenate WAMFs and WPMFs in order to simultaneously filter universal noise.  We focus on an explicit example providing first-order amplitude and dephasing noise filtering.  The basic procedure is to implement each constant-amplitude segment of a four-segment WAMF, $W1$, using a constant-amplitude phase-modulated sequence robust against amplitude noise. As a reminder, the noise-filtering performance of $W1$ is shown in Fig. ~\ref{Fig:Concatenation}c. Here we use the WPMF$\equiv$SK1 sequence for the phase modulation. We refer to the concatenated gate as a Universal Walsh Modulated Gate, UWMF$_{1,SK1}$.

 Referring to Eq. \ref{WAGMOrder1CPSMatrix},  the WAMF filter is similarly written $P_3(\WAMGRabiPlus/4,0)P_2(\WAMGRabiMin/2,0)P_1(\WAMGRabiPlus/4,0)$. Concatenation then involves the operator substitutions
\begin{align}
P_1(\WAMGRabiPlus/4,0) &\rightarrow \text{SK1}^{(1)}(\WAMGRabiPlus/4)\\
P_2(\WAMGRabiMin/2,0) &\rightarrow  \text{SK1}^{(2)}(\WAMGRabiMin/2)\\
P_3(\WAMGRabiPlus/4,0) &\rightarrow \text{SK1}^{(3)}(\WAMGRabiPlus/4).
\end{align}
The composite structure for UWMF$_{1,SK1}$ is shown in Fig. ~\ref{Fig:Concatenation}a. Here the SK1 phase flips $\phi = \pm\phiSK$ within each segment of the WAMF profile are indicated by the oppositely oriented hatching; $\phi = 0$ is indicated by white fill. The dephasing and amplitude filter functions for the concatenated sequence are shown in Fig. ~\ref{Fig:Concatenation}d, indicating effective filtering of both amplitude and dephasing noise.

\subsection{Ytterbium Ion Trapping}
We use trapped \textsuperscript{171}Yb\textsuperscript{+} ions as our experimental platform; a detailed description of related experimental approaches appears in~\cite{OlmschenkPRA2007, SoareBath}.  A linear Paul trap enclosed in an ultra-high vacuum (UHV) chamber is used to trap several hundred \textsuperscript{171}Yb\textsuperscript{+} ions as a small homogeneous ensemble (in magnetic field and microwave field amplitude).  Doppler cooling of the ions is achieved using 369 nm laser light, slightly red-detuned from the \textsuperscript{2}S$_{1/2}$ to \textsuperscript{2}P$_{1/2}$ transition.   Additional lasers near 935 nm and 638 nm are employed to depopulate metastable states. 

Our qubit is the 12.6 GHz hyperfine splitting between the \textsuperscript{2}S$_{1/2}\ket{F=0,m_{f}=0}$ and \textsuperscript{2}S$_{1/2}\ket{F=1,m_{f}=0}$ states. For notational simplicity we will designate $\ket{0}$ and $\ket{1}$ to these states respectively.   Addition of a 2.1 GHz sideband to the 369nm laser using an electro-optic modulator permits high-fidelity state preparation in $\ket{0}$.  For details of ion loading, laser cooling, state preparation, and state detection see~\cite{SoareBath}.  While we typically employ a small ensemble of ions, the system behaves similarly to single-ion experiments in our lab, and benefits from both high-fidelity state initialization and projective measurement - the system does not bear similarity to NMR-style ensembles.  

State detection is achieved by counting 369 nm photons scattered from the ions and converting to a probability that the Bloch vector lies at a particular location along a meridian of the Bloch sphere. This measurement is susceptible ion loss in the ensemble and both laser amplitude and frequency drifts over long timescales, resulting in variable maximum and dark count rates over time.  We therefore employ a normalization and Bayesian estimation procedure for state detection, see~\cite{SoareBath}.

An important advantage of this system is that the selected qubit transition is first order insensitive to magnetic field fluctuations; the measured free-evolution in our setup is $T_{2}\approx4$ s, limited by coherence between the qubit and the master oscillator~\cite{SoareBath}.  Coherent rotations between the measurement basis states are driven by using the magnetic field component of resonant microwave radiation.  The Rabi rate for driven oscillations reaches $\sim14\;\mu$s in our system, with typical operation near $\sim50\;\mu$s.  Rotations are implemented about an axis $\vec{r}$ lying on the $xy$-plane of the Bloch sphere and set by the phase of the microwaves as $\vec{r}=\left(\cos\phi(t), \sin\phi(t),0\right)$.  Driven operations, characterized by randomized benchmarking, exhibit a mean fidelity in excess of $99.99\%$.

\subsubsection{Noise Engineering}
In the laboratory we rely on engineering noise in our control system to provide a method to accurately reproduce decoherence processes of interest.  We begin with a desired noise power spectral density in either the amplitude or detuning quadrature (or both), assuming they are statistically independent.  From this power spectrum, defined by the noise strength $\alpha$, the exponent of the power-law scaling $p$, the comb spacing $\omega_{0}$, and the high-frequency cutoff $\omega_{c}\geq J\omega_{0}$, we numerically generate time-domain vectors for amplitude and frequency errors.  Noise is injected into the system by adding these modulation patterns on top of the control sequence being implemented (e.g. a pulse of radiation for implementing a $\pi$-pulse)using $IQ$ modulation in our vector signal generator~\cite{SoareBath}. 

\subsubsection{Randomized Benchmarking}
We use randomized benchmarking as a tool for resolving small gate errors which cannot be resolved in the application of a single gate. Our randomized benchmarking sequence consists of interleaved $\pi/2$ and $\pi$ pulses each applied along axes randomly selected from $\pm x$ and $\pm y$. Each sequential pair of $\pi/2$ and $\pi$ rotations is referred to a computational gate. A given randomized benchmarking sequence consists of $l$ computational gates followed by a final correcting gate which is selected such that the aggregate Unitary operation applied is a $\pi$ rotation.  For each $l$ we measure 50 randomizations (dots in Fig.~\ref{Fig:F2}f), and in each randomization average over 20 different realizations of a white dephasing noise bath.  Each realization, in turn, employs 20 measurements in our Bayesian state-detection algorithm, in addition to associated normalization experiments.

Comparisons of $W1$ to primitive $\pi$ rotation performance in randomized benchmarking is conducted via replacement of all $\pi$ pulses with $W1$ constructions, again about randomly selected axes.   In either case the $\pi/2$ rotation is achieved using a primitive gate, although we have also validated that replacement of the $\pi/2$ gates with WAMF constructions yields net improvement in gate fidelity.

\end{document}